\documentclass[prc,twocolumn,twosidenofootinbib,floatfix]{revtex4}

\usepackage{graphicx,color}
\usepackage{amsmath,amssymb,bm}
\usepackage{times}

\newcommand{\eq}[1]{\begin{equation}#1\end{equation}}
\newcommand{\eqmulti}[1]{\begin{equation}\begin{split}#1\end{split}\end{equation}}

\newcommand{\bra}[1]{\ensuremath{\langle{#1}|\,}}

\newcommand{\ket}[1]{\ensuremath{\,|{#1}\rangle}}

\newcommand{\braket}[2]{\ensuremath{\langle{#1}|{#2}\rangle}}

\newcommand{\matrixe}[3]{\ensuremath{\langle{#1}|\,{#2}\,|{#3}\rangle}}

\newcommand{\comm}[2]{\ensuremath{[{#1},{#2}]}}

\newcommand{\op}[1]{\ensuremath{#1}}
\newcommand{\adj}[1]{\ensuremath{{{#1}}^{\dag}}}
\newcommand{\corr}[1]{\ensuremath{\widetilde{#1}}}

\renewcommand{\vec}[1]{\ensuremath{\bm{#1}}}

\newcommand{\gO}{\ensuremath{\op{g}}}

\newcommand{\qO}{\ensuremath{\op{q}}}
\newcommand{\rO}{\ensuremath{\op{r}}}

\newcommand{\etaO}{\ensuremath{\op{\eta}}}

\newcommand{\CO}{\ensuremath{\op{C}}}

\newcommand{\HO}{\ensuremath{\op{H}}}

\newcommand{\OO}{\ensuremath{\op{O}}}

\newcommand{\TO}{\ensuremath{\op{T}}}
\newcommand{\UO}{\ensuremath{\op{U}}}
\newcommand{\VO}{\ensuremath{\op{V}}}

\newcommand{\CCO}{\ensuremath{\adj{\op{C}}}}
\newcommand{\UUO}{\ensuremath{\adj{\op{U}}}}

\newcommand{\idO}{\ensuremath{\op{1}}}

\newcommand{\qOV}{\ensuremath{\vec{\op{q}}}}
\newcommand{\rOV}{\ensuremath{\vec{\op{r}}}}

\newcommand{\LOV}{\ensuremath{\vec{\op{L}}}}

\newcommand{\sigmaOV}{\ensuremath{\vec{\op{\sigma}}}}

\newcommand{\Trel}{\ensuremath{\TO_\text{rel}}}

\newcommand{\tautauO}{\ensuremath{(\vec{\op{\tau}}_{\!1}\!\cdot\!\vec{\op{\tau}}_{\!2})}}
\newcommand{\sigmasigmaO}{\ensuremath{(\vec{\op{\sigma}}_{\!1}\!\cdot\!\vec{\op{\sigma}}_{\!2})}}

\newcommand{\tensorRRO}{\ensuremath{\op{S}_{12}(\tfrac{\rOV}{\rO},\tfrac{\rOV}{\rO})}}

\newcommand{\tensorRQO}{\ensuremath{\op{S}_{12}(\rOV,\qOV_{\Omega})}}

\newcommand{\spinorbitO}{\ensuremath{(\vec{\op{L}}\cdot\vec{\op{S}})}}

\newcommand{\Rm}{\ensuremath{R_-}}
\newcommand{\DRm}{\ensuremath{R'_-}}

\newcommand{\Rp}{\ensuremath{R_+}}

\newcommand{\Rpm}{\ensuremath{R_{\pm}}}

\newcommand{\UCOM}{\ensuremath{\textrm{UCOM}}}

\newcommand{\cm}{\ensuremath{\textrm{cm}}}
\newcommand{\rel}{\ensuremath{\textrm{rel}}}
\newcommand{\elem}[2]{\ensuremath{{}^{#2}\text{#1}}}

\newcommand{\fm}{\ensuremath{\,\text{fm}}}

\newcommand{\MeV}{\ensuremath{\,\text{MeV}}}

\newcommand{\symboldiamond}[1][black]{{\color{#1}$\blacklozenge$}}

\newcommand{\symbolbox}[1][black]{{\color{#1}$\blacksquare$}}
\newcommand{\symbolcircle}[1][black]{{\color{#1}$\bullet$}}

\definecolor{FGViolet}{rgb}{0.61,0.32,0.61}
\definecolor{FGDarkBlue}{rgb}{0,0,0.6}
\definecolor{FGBlue}{rgb}{0,0,0.8}
\definecolor{FGLightBlue}{rgb}{0.2, 0.6, 0.8}
\definecolor{FGGreen}{rgb}{0.2,0.7,0.2}
\definecolor{FGLightGreen}{rgb}{0.4,1,0.4}
\definecolor{FGYellow}{rgb}{1,0.95,0}
\definecolor{FGOrange}{rgb}{0.95,0.5,0.1}
\definecolor{FGRed}{rgb}{0.8,0,0}
\definecolor{FGWhite}{rgb}{1,1,1}
\definecolor{FGLightGray}{rgb}{0.8,0.8,0.8}
\definecolor{FGGray}{rgb}{0.5,0.5,0.5}
\definecolor{FGDarkGray}{rgb}{0.3,0.3,0.3}
\definecolor{FGBlack}{rgb}{0,0,0}

\newcommand{\linethinsolid}[1][black]{\unitlength 0.7ex
  {\color{#1}
  \begin{picture}(6,1)
  \linethickness{0.18mm}
  \put(0,0.9){\line(1,0){6.0}}
  \end{picture}}\nolinebreak
}

\newcommand{\linethindashed}[1][black]{\unitlength 0.7ex
  {\color{#1}
  \begin{picture}(6,1)
  \linethickness{0.18mm}
  \put(0,0.9){\line(1,0){1.4}}
  \put(2.2,0.9){\line(1,0){1.4}}
  \put(4.4,0.9){\line(1,0){1.4}}
  \end{picture}}\nolinebreak
}

\newcommand{\linethindotted}[1][black]{\unitlength 0.7ex
  {\color{#1}
  \begin{picture}(6,1)
  \linethickness{0.18mm}
  \put(0,0.9){\line(1,0){0.7}}
  \put(1.2,0.9){\line(1,0){0.7}}
  \put(2.4,0.9){\line(1,0){0.7}}
  \put(3.6,0.9){\line(1,0){0.7}}
  \put(4.8,0.9){\line(1,0){0.7}}
  \end{picture}}\nolinebreak
}

\newcommand{\linethindashdot}[1][black]{\unitlength 0.7ex
  {\color{#1}
  \begin{picture}(6,1)
  \linethickness{0.18mm}
  \put(0,0.9){\line(1,0){0.4}}
  \put(1.0,0.9){\line(1,0){1.3}}
  \put(2.9,0.9){\line(1,0){0.4}}
  \put(3.9,0.9){\line(1,0){1.3}}
  \put(5.8,0.9){\line(1,0){0.4}}
  \end{picture}}\nolinebreak
}

\begin{document}

\title{Unitary Correlation Operator Method and Similarity Renormalization Group: \\Connections and Differences}

\author{R. Roth}
\email{Robert.Roth@physik.tu-darmstadt.de}

\author{S. Reinhardt}

\author{H. Hergert}

\affiliation{Institut f\"ur Kernphysik, Technische Universit\"at Darmstadt,
64289 Darmstadt, Germany}

\date{\today}

\begin{abstract}    

We discuss relations and differences between two methods for the construction of unitarily transformed effective interactions, the Similarity Renormalization Group (SRG) and Unitary Correlation Operator Method (UCOM). The aim of both methods is to construct a soft phase-shift equivalent effective interaction which is well suited for many-body calculations in limited model spaces. After contrasting the two conceptual frameworks, we establish a formal connection between the initial SRG-generator and the static generators of the UCOM transformation. Furthermore we propose a mapping procedure to extract UCOM correlation functions from the SRG evolution. We compare the effective interactions resulting from the UCOM-transformation and  the SRG-evolution on the level of matrix elements, in no-core shell model calculations of light nuclei, and in Hartree-Fock calculations up to $^{208}$Pb. Both interactions exhibit very similar convergence properties in light nuclei but show a different systematic behavior as function of particle number. 

\end{abstract}


\maketitle


\section{Introduction}

In the past few years several methods for the construction of phase-shift equivalent soft interactions starting from modern realistic potential have been proposed and applied. The common goal of these methods is to adapt realistic QCD-motivated interactions like the ones extracted from chiral effective field theory \cite{EpNo02,EnMa03} or more phenomenological high-precision potentials like the Argonne V18 \cite{WiSt95} to the limited model spaces typically available in many-body calculations. Apart from approaches providing effective interactions tailored for a specific model space, e.g. the Lee-Suzuki transformation \cite{SuLe80} widely used in the ab initio no-core shell model \cite{NaVa00,NaOr02}, there are several schemes to derive model-space independent effective interactions, e.g. the $V_{\text{low}k}$ approach providing a universal low-momentum interaction \cite{BoKu03}. 

We focus on two alternative schemes, the Unitary Correlation Operator Method (UCOM) \cite{FeNe98,NeFe03,RoNe04} and the Similarity Renormalization Group (SRG) \cite{Wegn94,Wegn00}, which both use phase-shift conserving unitary transformations. The physical picture behind these two formulations is different: The UCOM starts out from a coordinate-space representation of the short-range correlations induced by the central and tensor components of the realistic nuclear interaction. On this basis ansatzes for the generators of unitary transformations describing central and tensor correlations are formulated which allow for the explicit inclusion of these correlations in simple model spaces. The SRG, on the other hand, aims at the pre-diagonalization of a matrix representation of the Hamiltonian in a chosen basis by means of a renormalization group flow evolution. The resulting band-diagonal interaction is also well-suited for small model spaces. 

We discuss the formal relations and the practical differences between these two approaches in detail. After reviewing the formalism of the SRG and discussing examples for the evolution of matrix elements and two-body wavefunctions in Sec. \ref{sec:srg}, we put the UCOM approach in perspective. In Sec. \ref{sec:ucom} we discuss the formal connections between SRG and the generators of the UCOM transformation. Based on these structural relations, we propose a mapping scheme to extract UCOM correlation functions from the SRG evolution in Sec. \ref{sec:srgcorr}. Following a comparison of matrix elements of the SRG and UCOM-transformed interactions, we compare their behavior in different many-body calculations. In Sec. \ref{sec:ncsmhf} we present no-core shell model calculations for \elem{H}{3} and \elem{He}{4} and Hartree-Fock calculations for closed shell nuclei up to \elem{Pb}{208} for the different transformed interactions.      

\section{Similarity Renormalization Group}
\label{sec:srg}

\subsection{Concept \& Formalism}

The basic idea of the Similarity Renormalization Group (SRG) approach in the formulation of Wegner \cite{Wegn94,Wegn00} is to transform the initial Hamiltonian $\HO$ of a many-body system into a diagonal form with respect to a given basis. The renormalization group flow equation governing the evolution of the Hamiltonian is of the form 
\eq{ \label{eq:srg_flow}
  \frac{d\HO_\alpha}{d\alpha}
  = \comm{\etaO_\alpha}{\HO_\alpha} \;,
}
where $\alpha$ is the flow parameter and $\HO_{\alpha}$ the evolved Hamiltonian with $\HO_0=\HO$. Analogous equations can be formulated for the operators of all observables one is interested in. In general terms the anti-hermitian generator $\etaO_\alpha$ of the flow can be written as
\eq{ \label{eq:srg_generator0}
  \etaO_\alpha 
  = \comm{\text{diag}(\HO_\alpha)}{\HO_\alpha} \;,
}
where $\text{diag}(\HO_\alpha)$ refers to the diagonal part of the Hamiltonian in a given basis. This choice can be understood in intuitive terms: if the Hamiltonian commutes with its diagonal part w.r.t. a given basis, then the generator vanishes and the evolution has reached a fix point. Apart from trivial cases this does only happen if the Hamiltonian is actually diagonal in the given basis.

Formally one can integrate this flow equation defining a unitary operator $\UO_\alpha$ of the transformation   
\eq{ \label{eq:srg_unitrafo}
  \HO_\alpha
  = \UO_\alpha\HO\UUO_\alpha \;.
}
Due to the nontrivial $\alpha$-dependence of the generator, the unitary operator is not simply given by an exponential of the generator. Nevertheless, from \eqref{eq:srg_unitrafo} and \eqref{eq:srg_flow} we can construct a differential equation for the operator $\UO_\alpha$, 
\eq{ \label{eq:srg_unidgl}
  \frac{d\UO_\alpha}{d\alpha}
  = \etaO_\alpha \UO_\alpha \;,
}
with the initial condition $\UO_0 = \idO$, whose formal solution can be written as a Dyson series. Hence, for a given generator $\etaO_\alpha$ one either has to solve the flow equation \eqref{eq:srg_flow} for all operators of interest or one determines the unitary operator via \eqref{eq:srg_unidgl} and transforms all operators via \eqref{eq:srg_unitrafo}. 

So far this concept is generic and independent of the properties of the particular physical system, the Hamiltonian, or the basis under consideration. If considering an $A$-body system, then all the aforementioned relations refer to the operators in $A$-body space. One of the consequences is that even a simple initial Hamiltonian, containing two-body operators at most, acquires up to $A$-body terms in the course of the evolution. For practical applications of the SRG approach in the nuclear structure context one therefore has to simplify the scheme by confining the evolution to two or three-body space, thus discarding higher-order contributions in the evolved interaction. Furthermore, instead of using the diagonal part of the Hamiltonian in the definition of the generator, one can use the operator that defines the eigenbasis with respect to which the Hamiltonian shall be diagonalized. 

A simplified scheme suggested by Szpigel and Perry \cite{SzPe00} and applied by Bogner et al. \cite{BoFu07,HeRo07} confines the evolution to two-body space and uses the generator   
\eq{ \label{eq:srg_generator}
  \etaO_\alpha=\comm{\Trel}{\HO_\alpha}=\comm{\frac{\qOV^2}{2\mu}}{\HO_\alpha}\,,
}
containing the relative kinetic energy $\Trel = \frac{1}{2\mu}\qOV^2$ in the two-body system. The square of the two-body relative momentum operator can be decomposed into a radial and an angular part, 
\eq{ \label{eq:srg_momentumop}
  \qOV^2 
  = \qO_r^2 + \frac{\LOV^2}{\rO^2}
  \;,\qquad
  \qO_r 
  = \frac{1}{2}\Big(\qOV\cdot\frac{\rOV}{r} + \frac{\rOV}{r}\cdot\qOV\Big) \;.
}
The obvious fix point of the evolution with the explicit generator \eqref{eq:srg_generator} is a two-body Hamiltonian $\HO_\alpha$ that commutes with $\qO_r^2$ and $\LOV^2/\rO^2$. Hence, in a partial-wave momentum-space basis $\ket{q(LS)JT}$ this generator drives the matrix elements $\matrixe{q(LS)JT}{\HO_{\alpha}}{q'(L'S)JT}$ towards a band-diagonal structure with respect to relative momentum $(q,q')$ and orbital angular momentum $(L,L')$. Though we will only use this generator in the following, one should note that there are other physically motivated choices for $\etaO_\alpha$. An evident alternative for the operator $\qOV^2$ is the single-particle Hamiltonian of the harmonic oscillator.

Starting from an initial two-body Hamiltonian $\HO$ composed of relative kinetic energy $\TO_\rel$ and two-body interaction $\VO$ it is convenient to decompose the SRG-evolved Hamiltonian $\HO_\alpha$ in a similar way
\eq{ \label{eq:srg_definitiov}
  \HO_{\alpha} = \TO_{\text{rel}} + \VO_{\alpha} \;.
}
All flow-dependence is absorbed in the SRG-evolved two-body interaction $\VO_\alpha$ defined by this relation. Rewriting of the flow equation \eqref{eq:srg_flow} using the generator \eqref{eq:srg_generator} explicitly for the evolved interaction $\VO_\alpha$ leads to 
\eq{ \label{eq:srg_flowv}
  \frac{d\VO_\alpha}{d\alpha}
  = \comm{\etaO_\alpha}{\TO_{\text{rel}} + \VO_\alpha} 
  = \comm{ \comm{\TO_\rel}{\VO_\alpha}}{\TO_{\text{rel}} + \VO_\alpha}  \;.
}

Even in this simplified form a direct solution of the operator equation is far from trivial. For practical applications we therefore resort to the level of matrix elements. Given the ansatz \eqref{eq:srg_generator} for the generator, it is convenient to work in momentum space. 
Using the partial-wave momentum-space basis $\ket{q(LS)JT}$ the flow equation \eqref{eq:srg_flowv} translates into a set of coupled integro-differential equations for the matrix elements 
\eq{
  V^{(JLL'ST)}_{\alpha}(q,q')
  = \matrixe{q(LS)JT}{\VO_{\alpha}}{q'(L'S)JT} \;,
}
where the projection quantum numbers $M$ and $M_T$ have been omitted for brevity. In a generic form, the resulting evolution equation reads: 
\eqmulti{ \label{eq:srg_flowvme}
  &\frac{d}{d\alpha} V_{\alpha}(q,q') \\
  &= -\frac{1}{(2\mu)^2} (q^2-q'^2)^2\; V_{\alpha}(q,q') \\
  &\quad+\frac{1}{2\mu} \int dQ\, Q^2\; (q^2 + q'^2 -2 Q^2)\; 
    V_{\alpha}(q,Q) V_{\alpha}(Q,q') \;.
}
For non-coupled partial waves with $L=L'=J$, the matrix elements entering into this equation are simply 
\eq{ \label{eq:srg_vme_noncoupled}
  V_{\alpha}(q,q') 
  = V^{(JJJST)}_{\alpha}(q,q') \;.
}
For coupled partial waves with $L,L'=J\pm1$, the $V_{\alpha}(q,q')$ are understood as $2\times2$ matrices of the matrix elements for the different combinations of the orbital angular momenta $L = J-1$ and $L'=J+1$
\eq{ \label{eq:srg_vme_coupled}
  V_{\alpha}(q,q') 
  = \begin{pmatrix}
    V^{(JLLST)}_{\alpha}(q,q')  & V^{(JLL'ST)}_{\alpha}(q,q') \\
    V^{(JL'LST)}_{\alpha}(q,q')  & V^{(JL'L'ST)}_{\alpha}(q,q') \\
  \end{pmatrix} \;.
}
Each non-coupled partial wave and each set of coupled partial waves evolves independently of the other channels of the interaction. This is a direct consequence of the choice of the generator---the evolution towards a diagonal in momentum space is done in an optimal way for each individual partial wave.

As mentioned earlier, analogous evolution equations have to be solved for all observables in order to arrive at a consistent set of effective operators. The evolution of these operators, e.g. the multipole operators necessary for the evaluation of transition strengths or the one-body density operators employed for the computation of the momentum distribution, is coupled to the evolution of the Hamiltonian via the generator $\eta_\alpha$. Hence we have to solve these evolution equations simultaneously. 

An alternative approach is to determine the matrix elements of the unitary operator $\UO_\alpha$ explicitly by solving \eqref{eq:srg_unidgl}. The evolved matrix elements of all observables can then be obtained by a simple matrix transformation using the same unitary transformation matrix. In the case of the momentum-space partial-wave matrix elements of the unitary transformation operator,
\eq{
  U^{(JLL'ST)}_{\alpha}(q,q')
  = \matrixe{q(LS)JT}{\UO_{\alpha}}{q'(L'S)JT} \;,
}
the operator equation \eqref{eq:srg_unidgl} leads to a coupled set of integro-differential equations 
\eqmulti{ \label{eq:srg_flowume}
  &\frac{d}{d\alpha} U_{\alpha}(q,q') \\
  &= \frac{1}{2\mu} \int dQ\, Q^2\; (q^2 - Q^2)\; 
    V_{\alpha}(q,Q) U_{\alpha}(Q,q') \;,
}
where we assume that the evolution equation \eqref{eq:srg_flowvme} is solved simultaneously providing the $V_{\alpha}(q,q')$. The generic notation defined in \eqref{eq:srg_vme_noncoupled} and \eqref{eq:srg_vme_coupled} for non-coupled and coupled partial waves, respectively, applies here as well. This differential equation provides direct access to the matrix elements of the unitary operator, which maps the initial operators onto any particular point of the flow trajectory. 

Note that the concept of the SRG discussed so far is independent of the particular physical system and the properties of the Hamiltonian under consideration. The only restriction concerns the basis with respect to which the Hamiltonian shall be diagonalized. This is different from the motivation of the Unitary Correlation Operator Method discussed in the Sec. \ref{sec:ucom}.

\subsection{Numerical Examples}
\label{sec:srg_illu}

\begin{figure*}
\includegraphics[width=0.276\textwidth]{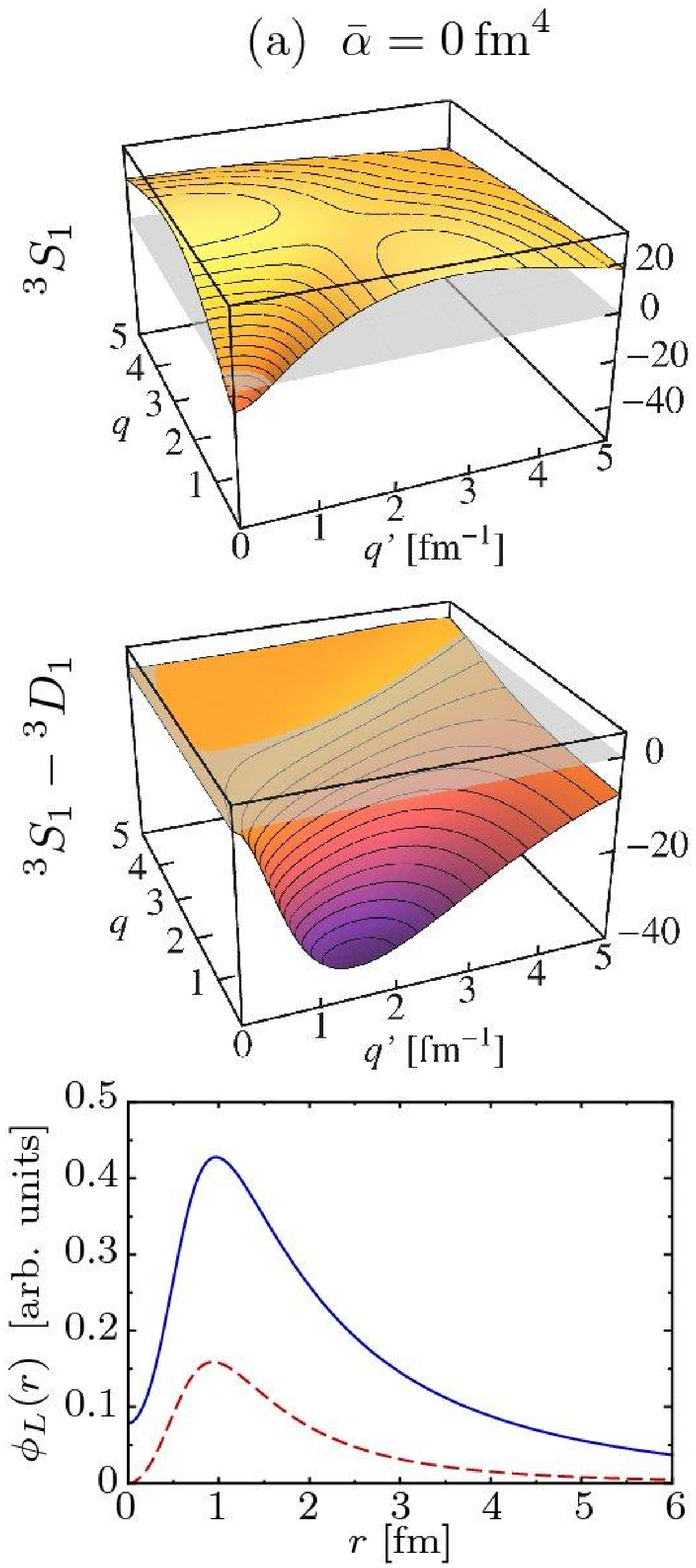}%
\includegraphics[width=0.24\textwidth]{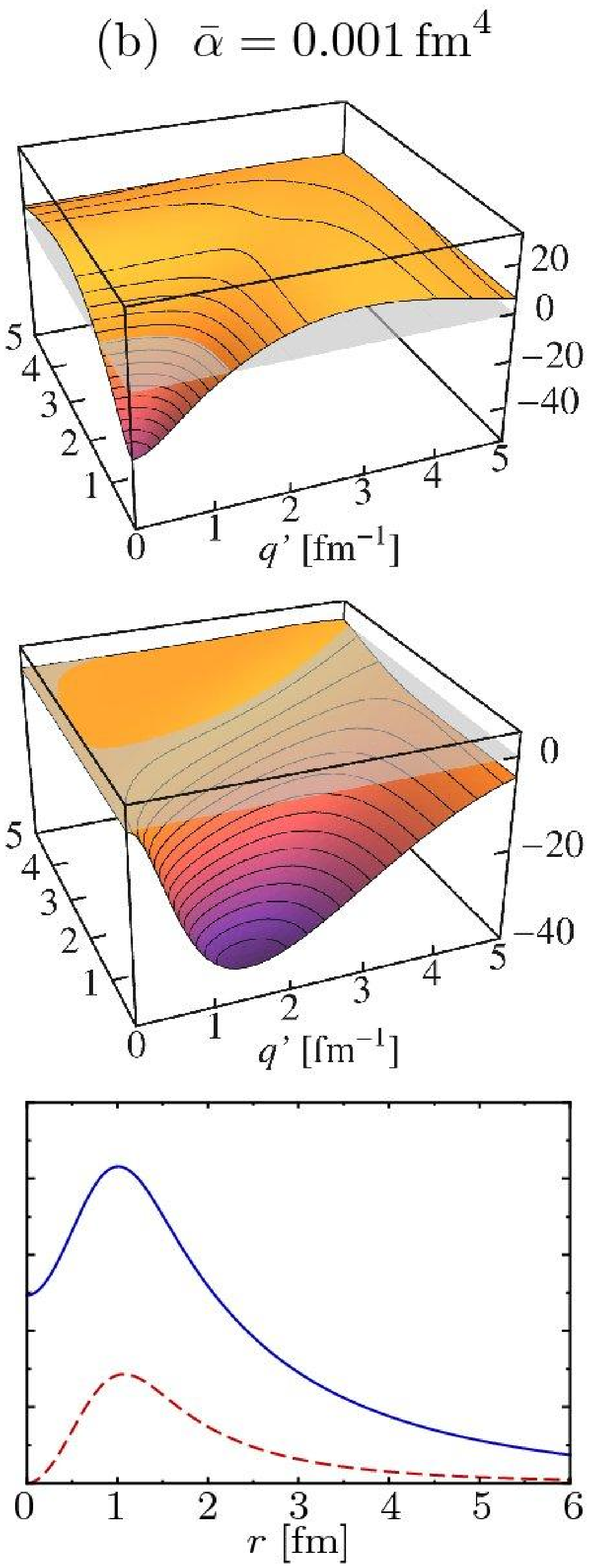}%
\includegraphics[width=0.24\textwidth]{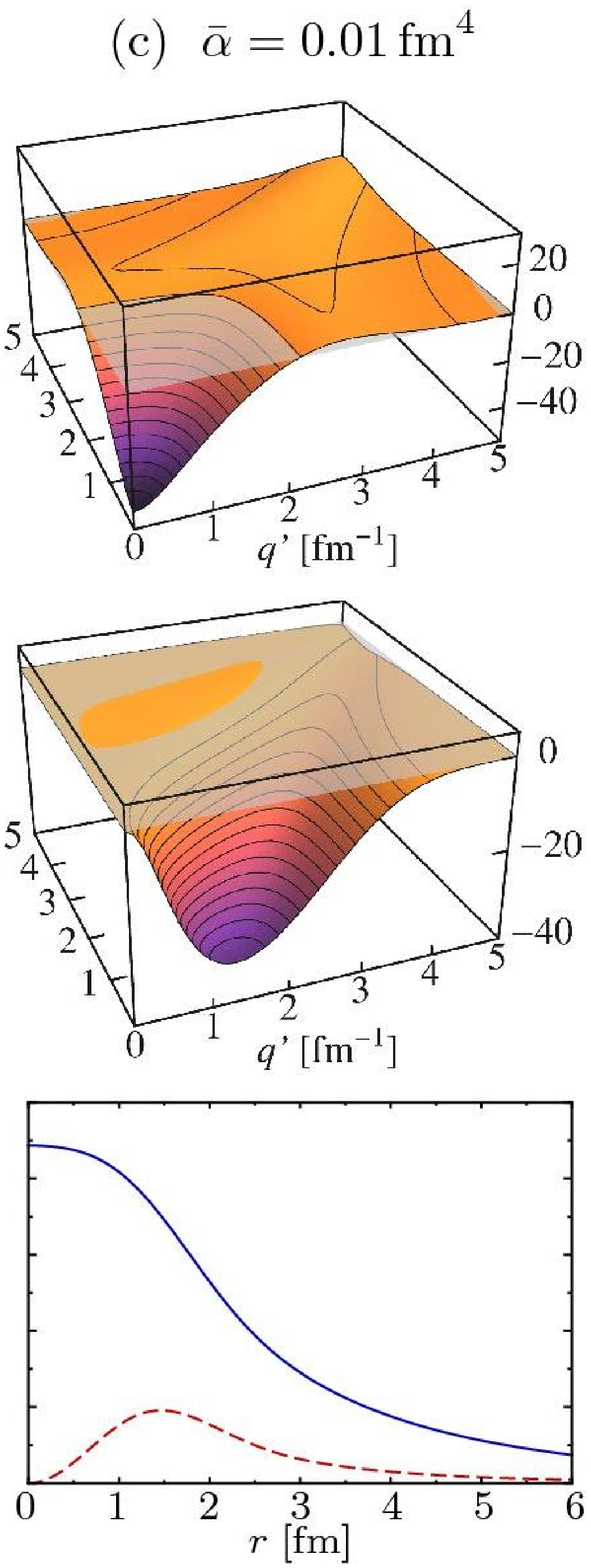}%
\includegraphics[width=0.24\textwidth]{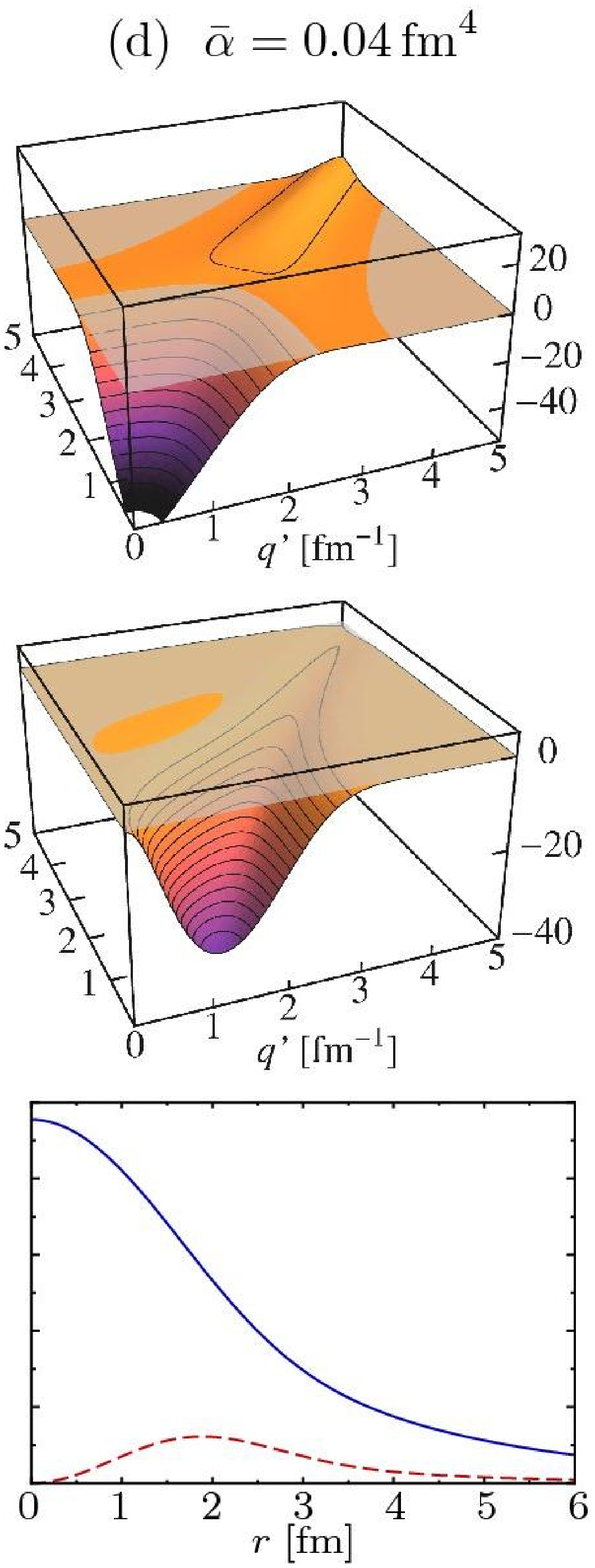}
\caption{(color online) Illustration of the SRG evolution of the momentum-space matrix elements and the deuteron wavefunction starting from the Argonne V18 potential. The four columns correspond to different values of the flow parameter: (a) $\bar{\alpha}=0\,\fm^4$, (b) $0.001\,\fm^4$, (c) $0.01\,\fm^4$, and (d) $0.04\,\fm^4$. The upper two rows depict the matrix elements $V_\alpha(q,q')$ for the $^3S_1$ and the $^3S_1-^3D_1$ partial waves, resp., in units of $\MeV\fm^3$. The bottom row shows the radial coordinate-space wavefunctions $\phi_L(r)$ of the deuteron ground state obtained with the respective SRG-evolved matrix elements: (\linethinsolid[FGBlue]) $L=0$, (\linethindashed[FGRed]) $L=2$.}
\label{fig:srg_illu}
\end{figure*}

In order to illustrate the impact of the flow evolution on the properties of the SRG interaction, we discuss the momentum-space matrix elements for selected partial waves as well as the deuteron solution obtained with these matrix elements. We start out from the Argonne V18 (AV18) potential \cite{WiSt95} as an example for interactions with a strongly repulsive core.

The numerical solution of the evolution equations \eqref{eq:srg_flowvme} and \eqref{eq:srg_flowume} for the matrix elements is straightforward. For convenience we absorb the mass factors into a rescaled flow parameter $\bar{\alpha} = \alpha/(2\mu)^2$ given in units of $\fm^4$ and rescaled interaction matrix elements $\bar{V}_{\alpha}(q,q') = (2\mu)\,V_{\alpha}(q,q')$ given in units of $\fm$. After discretizing the momentum variables, the coupled set of first-order differential equations can be solved with standard methods, e.g. an adaptive Runge-Kutta algorithm. The numerical solution is robust against changes of the discretization pattern. However, one has to make sure that the momentum range covered by the grid is sufficiently large such that the matrix elements of the initial potential are zero at and beyond the boundaries. 

Based on the SRG evolved momentum-space matrix elements, we can solve the two-body problem in a given partial wave in momentum space using the same discretized momentum grid as before. In addition to the bound deuteron solution we obtain discretized continuum states resulting from the boundary conditions. The eigenvalues of the two-body problem provide an additional check for the accuracy of the numerical scheme. Since in two-body space the evolution equations \eqref{eq:srg_flowvme} correspond to the complete unitary transformation, the spectrum of the Hamiltonian is preserved. All energy eigenvalues resulting for any two-body system have to be independent of $\alpha$. In our numerical calculation this is fulfilled to an relative accuracy of better than $10^{-6}$. Of course, the eigenstates obtained for the two-body system do depend on the flow parameter $\alpha$. After transformation to coordinate space, the resulting wavefunctions provide a direct illustration of the interaction-induced correlations and their reduction throughout the SRG evolution.

Figure \ref{fig:srg_illu} illustrates the effect of the SRG evolution in the deuteron channel. The 3D plots in the upper two rows show the momentum-space matrix elements for the $^3S_1$ and $^3S_1-{}^3D_1$ partial waves. The plots in the lower row depict the $S$- and $D$-wave component of the radial deuteron wavefunction in coordinate representation obtained from the solution of the two-body problem for the evolved interaction. Each column of plots corresponds to a different value of the flow parameter starting from $\bar{\alpha}=0\,\fm^4$, i.e. the initial AV18 potential, to $\bar{\alpha}=0.04\,\fm^4$, which is a typical value for the later applications. The matrix elements clearly show how the flow evolution drives the matrix towards band-diagonal structure. The initial AV18 matrix elements in the $^3S_1$ channel have large off-diagonal contributions ranging to very high momentum differences $|q-q'|$. In the course of the SRG evolution, the off-diagonal matrix elements are suppressed, any high-momentum components are concentrated along the diagonal, and the attractive low-momentum part is enhanced. Similarly, in the $^3S_1-{}^3D_1$ partial wave, the strong off-diagonal contributions caused by the tensor interaction are eliminated outside of a band along the diagonal. Analogous effects are observed in all other partial waves.

The impact of this pre-diagonalization in momentum space on the ground state wavefunction of the deuteron in coordinate representation (bottom row in Fig. \ref{fig:srg_illu}) is remarkable. The two distinct manifestations of short-range correlations---the suppression of the relative wavefunction at small interparticle distances (as a result of the short-range repulsion) and the presence of the $D$-wave admixture (as a result of the strong tensor interaction)---are gradually eliminated during the flow evolution. Already for very small flow parameters $\alpha$, the structures at the shortest distances, corresponding to large momenta, are removed. The transformed interaction $\VO_{\alpha}$ for $\bar{\alpha}\gtrsim0.01\,\fm^4$ does not generate a correlation hole in the wavefunction anymore. With increasing flow parameter, structures at larger and larger radii and thus smaller and smaller momenta are suppressed. In this way the $D$-wave admixture is systematically eliminated starting from small $r$. For the flow parameters $\bar{\alpha}\approx0.04\,\fm^4$ the $D$-wavefunction consists only of a weak contribution around $r\approx2\,\fm$. 
The quadrupole moment is conserved despite the elimination of the $D$-wave component, since the quadrupole operator itself has to be transformed and acquires a more complicated structure. Obviously, a consistent evolution of the Hamiltonian and all observables is mandatory. 

The simple example of the deuteron shows the connection between off-diagonal contributions of the interaction and correlations in coordinate space, providing a first link between the SRG and the Unitary Correlation Operator Method discussed in the following.

\section{Unitary Correlation Operator Method}
\label{sec:ucom}

\subsection{Concept \& Formalism}

The idea of the Unitary Correlation Operator Method (UCOM) \cite{RoHe05,RoNe04} is to include the most important short-range correlations induced by realistic nuclear interactions with an explicit unitary transformation described by a so-called correlation operator $\CO$. This unitary operator can be used to imprint the short-range correlations onto an uncorrelated many-body state $\ket{\Psi}$, leading to a correlated state 
\eq{
  \ket{\corr{\Psi}} = \CO \ket{\Psi} \;.
}
Alternatively, it can be used to define transformed or correlated operators for all observables of interest. The unitary transformation of the initial Hamiltonian $\HO$ leading to the correlated Hamiltonian $\corr{\HO}$ reads
\eq{
  \corr{\HO} = \CCO \HO \CO \;.
}

In contrast to the SRG approach, we choose an explicit ansatz for the unitary correlation operator $\CO$ which is motivated by physical considerations on the structure of the correlations induced by realistic nuclear interactions. First of all, we distinguish the correlations caused by the short-range repulsion in the central part of the interaction---so-called central correlations---and those induced by the tensor part---so-called tensor correlations. The correlation operator is written as a product of two unitary operators $\CO_\Omega$ and $\CO_r$ accounting for tensor and central correlations, respectively, each formulated via an exponential ansatz 
\eq{
   \CO 
   = \CO_\Omega \CO_r 
   = \exp\!\Big(\!-i\sum_{j<k}\gO_{\Omega,jk}\Big) \exp\!\Big(\!-i\sum_{j<k}\gO_{r,jk}\Big) \;.
}
Explicit expressions for the hermitian generators $\gO_{\Omega}$ and $\gO_{r}$ are constructed based on the physical mechanism responsible for the correlations, as discussed in detail in Refs. \cite{RoNe04,RoHe05}. 

The central correlations induced by the short-range respulsion are revealed through the suppression of the two-body density at short distances. Pictorially speaking, the interaction pushes close-by nucleons apart and thus out of the region of the mutual repulsion. This kind of distance-dependent radial shift is described by the generator 
\eq{ \label{eq:ucom_gen_r}
  \gO_r 
   = \frac{1}{2} \big(\qO_r s(r) + s(r)\qO_r \big)
}
with the radial component of the relative momentum operator defined in \eqref{eq:srg_momentumop}. The tensor correlations caused by the tensor part of the interaction connect the spin and angular degrees of freedom and result in the mixing of states with orbital angular momentum $L$ and $L\pm2$. This can be created with the generator
\eqmulti{ \label{eq:ucom_gen_tens}
  \gO_\Omega 
  &= \vartheta(r)\tensorRQO \\
  &= \vartheta(r)\frac{3}{2} 
     \big((\sigmaOV_1\cdot\qOV_{\Omega}) (\sigmaOV_2\cdot\rOV)
     + (\sigmaOV_1\cdot\rOV)(\sigmaOV_2\cdot\qOV_{\Omega}) \big) \,,
}
where $\qOV_{\Omega} = \qOV - \qO_r\frac{\rOV}{r}$. The strengths and distance-dependences of the two transformations are described by the functions $s(r)$ and $\vartheta(r)$ that depend on the potential under consideration. In general we will amend the generators by projection operators on two-body spin $S$ and isospin $T$ in order to allow for a spin-isospin dependence of the unitary transformation.

As for the SRG transformation, the correlated Hamiltonian will contain irreducible contributions to all particle numbers up to $A$ even if the initial Hamiltonian contains only two-body terms. Analogously, we restrict the discussion to two-body space thus discarding any higher-order contributions of the cluster expansion of correlated operators. Again, we decompose the correlated Hamiltonian $\corr{\HO}$ into the relative kinetic energy $\TO_\rel$ and the correlated interaction $\VO_\UCOM$ [cf. Eq. \eqref{eq:srg_definitiov}]
\eq{
  \corr{\HO} 
  = \CCO\HO\CO 
  = \TO_{\rel} + \VO_\UCOM   \;.
}
As one of the benefits of the explicit formulation of the correlation operator, we can derive an explicit operator form of the correlated interaction $\VO_\UCOM$ as well as analytic expressions for the transformed matrix elements. We will not discuss these aspects in detail but refer to Refs. \cite{RoHe05, RoNe04, NeFe03, FeNe98}.

\subsection{UCOM from an SRG Perspective}

One can consider the UCOM generators \eqref{eq:ucom_gen_r} and \eqref{eq:ucom_gen_tens} also from an SRG perspective. Though the aforementioned physical picture originally gave rise to the formulation of the UCOM \cite{FeNe98,NeFe03,RoNe04}, one can obtain the same operator structures in the framework of the SRG \cite{HeRo07}.
We assume an initial interaction composed of central, spin-orbit and tensor part,
\eq{
  \VO=\sum_p v_p(\rO)\,\OO_p 
}
with $\OO_p \in \{1, \sigmasigmaO, \spinorbitO, \tensorRRO,... \} \otimes \{\idO, \tautauO\}$. By evaluating the commutator \eqref{eq:srg_generator} explicitly for $\alpha=0$ using this operator form we obtain  
\eq{ \label{eq:ucom_gen_srg}
  \etaO_0
  =  \frac{i}{2} \big(\qO_r S(\rO) + S(\rO)\qO_r \big) + i \Theta(\rO) \tensorRQO \;.
}
The operator-valued functions $S(\rO)$ and $\Theta(\rO)$ contain the radial dependencies of the different terms of the interaction
\eq{ \label{eq:ucom_corrfunc_srg}
  S(\rO) = -\frac{1}{\mu} \bigg(\sum_pv'_p(\rO)\OO_p \bigg)
  \;,\quad
  \Theta(\rO) \equiv -\frac{2}{\mu}\frac{v_t(\rO)}{\rO^2}  \,. 
}
Thus, the initial SRG generator has the same operator structure as the UCOM generators $\gO_r$ and $\gO_\Omega$ that were constructed based on the physical picture of central and tensor correlations \cite{HeRo07}. 

First of all, this formal connection shows that both approaches address the same physics of short-range correlations, although starting from quite different backgrounds. Moreover, it proves that the set of UCOM generators covers the most relevant terms. Although there are other operators appearing in the initial interaction, e.g. the spin-orbit operator, they do not require separate generators---their effect on the correlations is absorbed in the operator-valued function $S(\rO)$.   

At this point UCOM uses a simplified strategy. The correlation functions $s(r)$ and $\vartheta(r)$ are chosen to depend on spin $S$ and isospin $T$ only, they do not depend on orbital and total angular momentum. Formally one could drop this restriction and work with separate correlation functions for each partial wave and thus mimic the flexibility of the SRG generator. In practice this does not seem necessary or advantageous. 

Although there is the direct relation between UCOM and initial SRG generators, this does not allow us to identify the UCOM correlation functions directly. In the language of SRG, a single UCOM transformation encapsulates a whole flow evolution up to a certain flow parameter $\alpha$. In order to extract UCOM correlation functions, we therefore have to solve the flow equation with the dynamical SRG generator. The initial SRG generator alone does not provide this information.

\subsection{Correlated Two-Body States}
\label{sec:ucom_states}

One aspect of the UCOM formalism of relevance for the following is the behavior of two-body states under the unitary transformation. The action of the central correlator $\CO_r$ on the relative component $\ket{\Phi}$ of a two-body state $\ket{\Psi}= \ket{\Phi} \otimes \ket{\Phi_{\cm}}$ can be evaluated directly in coordinate representation 
\eq{ \label{eq:ucom_central_states}
  \matrixe{r}{\CO_r}{\Phi}
  = \sqrt{\DRm(r)} \frac{\Rm(r)}{r} \braket{\Rm(r)}{\Phi} \;,
}
where the correlation function $\Rm(r)$ as well as its inverse $\Rp(r)$ are connected to $s(r)$ by the integral equation
\eq{
  \int_{r}^{\Rpm(r)}\frac{d\xi}{s(\xi)} = \pm1\,.
}
Hence the application of $\CO_r$ corresponds to a simple coordinate transformation $r \mapsto \Rpm(r)$ with the transformation function $\Rpm(r)$. 

The action of the tensor correlator $\CO_{\Omega}$ on a relative two-body state with definite angular momentum $\ket{\Phi} = \ket{\phi (LS)JT}$ can also be evaluated explicitly \cite{NeFe03, RoNe04, RoHe05}. States with $L=J$ are invariant under transformation with the tensor correlation operator
\eq{ \label{eq:corr_tensor_states1}
  \CO_{\Omega} \ket{\phi (J S) J T}
  = \ket{\phi (J S) J T} \;.
}
States with $L=J\pm1$ acquire an admixture of a component with $L'=J\mp1$ with a modified radial dependence 
\eqmulti{ \label{eq:corr_tensor_states2}
  \CO_{\Omega} \ket{\phi (J\pm 1,1) J T}
  &= \cos\theta_J(\rO)\, \ket{\phi (J\pm 1,1) J T} \\
  &\mp\, \sin\theta_J(\rO)\, \ket{\phi (J\mp 1,1) J T} \;,
}
where
\eq{
  \theta_J(\rO) 
  = 3 \sqrt{J(J+1)}\; \vartheta(\rO) \;.
}   
One can easily combine these two transformations obtaining a closed expression for a correlated two-body state in coordinate representation.

\section{UCOM Correlators Derived from SRG Solutions}
\label{sec:srgcorr}

Based on the elements introduced in Secs. \ref{sec:srg} and \ref{sec:ucom} we now devise a scheme to extract UCOM correlation functions from an SRG evolution of a given initial interaction. So far, the UCOM correlation functions have been determined through a variational calculation in the two-body system using simple parametrizations of the functions $\Rp(r)$ and $\vartheta(r)$ \cite{RoHe05}. The use of the SRG as a tool to construct UCOM correlation function has several conceptual advantages as will be discussed later on. In order to avoid confusion, we note from the outset that the UCOM transformation using SRG-generated correlation functions is not equivalent to the SRG transformation. 

The scheme for the construction of SRG-generated UCOM correlation functions consists of three steps: (\emph{i}) We solve the SRG evolution equations for a given initial interaction up to a flow parameter $\alpha$, obtaining the momentum space matrix elements $V_\alpha(q,q')$ for a certain partial wave. (\emph{ii}) Using the evolved matrix elements the two-body problem is solved leading to a set of coordinate-space wavefunctions. (\emph{iii}) The UCOM correlation functions $s(r)$ and $\vartheta(r)$ are determined such that they map a selected two-body eigenstate of the SRG evolved interaction onto the corresponding two-body state of the initial interaction in the respective partial wave. The steps (\emph{i}) and (\emph{ii}) have already been illustrated for the deuteron channel in Sec. \ref{sec:srg_illu}. Step (\emph{iii}) is discussed in the following.
  
\subsection{Mapping Solution}

Consider two eigenstates $\ket{\Phi^{(0)}}$ and $\ket{\Phi^{(\alpha)}}$ with the same energy eigenvalue resulting from the solution of the two-body problem for the initial and the SRG-evolved potential, respectively, in a given coupled or non-coupled partial wave. We can define a UCOM correlation operator $\CO$ that maps the two states onto each other 
\eq{ \label{eq:srgcorr_definition}
  \ket{\Phi^{(0)}} 
  = \CO \ket{\Phi^{(\alpha)}}
  = \CO_\Omega \CO_r \ket{\Phi^{(\alpha)}} \;.
}
Based on this formal definition we can derive equations that determine the correlation functions $\Rm(r)$ and $\vartheta(r)$ that characterize the correlation operator.

For non-coupled partial waves with $L=J$ only the central correlator appears. With the two-body solutions 
\eqmulti{
  \ket{\Phi^{(0)}} 
  &= \ket{\phi^{(0)} (LS)JT} \\
  \ket{\Phi^{(\alpha)}} 
  &= \ket{\phi^{(\alpha)} (LS)JT}  
}
for the initial and the SRG-evolved interaction, respectively, we obtain from \eqref{eq:srgcorr_definition} and \eqref{eq:ucom_central_states} a relation connecting the known radial wavefunctions $\phi^{(0)}(r)$ and $\phi^{(\alpha)}(r)$ via a yet unknown correlation function $\Rm(r)$:
\eq{ \label{eq:srgcorr_noncoupledpw_pre}
  \phi^{(0)}(r) 
  = \frac{\Rm(r)}{r} \sqrt{\DRm(r)}\; \phi^{(\alpha)}(\Rm(r)) \;.
}
Here and in the following we assume real-valued wavefunctions. The relation \eqref{eq:srgcorr_noncoupledpw_pre} can be viewed as a differential equation for the correlation function $\Rm(r)$. After formal integration we arrive at an implicit integral equation for $\Rm(r)$
\eq{  \label{eq:srgcorr_noncoupledpw}
  [\Rm(r)]^3 
  = 3 \int_0^r d\xi\; \xi^2 \frac{[\phi^{(0)}(\xi)]^2}{[\phi^{(\alpha)}(\Rm(\xi))]^2} \;,
}
which can be solved easily in an iterative fashion. We end up with a discretized representation of the correlation function $\Rm(r)$ for the partial wave under consideration. By construction it maps a selected SRG-evolved two-body state onto the corresponding initial state. In general, $\Rm(r)$ will depend on the pair of states, e.g. the ground states or a pair of excited states, we have selected. We will show later on that this dependence is very weak.

For coupled partial waves with $L=J-1$ and $L'=J+1$ central and tensor correlators act simultaneously. Using the two-body eigenstates 
\eqmulti{
  \ket{\Phi^{(0)}} 
  &= \ket{\phi^{(0)}_{L} (LS)JT} + \ket{\phi^{(0)}_{L'} (L'S)JT} \\
  \ket{\Phi^{(\alpha)}} 
  &= \ket{\phi^{(\alpha)}_{L} (LS)JT} + \ket{\phi^{(\alpha)}_{L'} (L'S)JT} 
}
of the initial interaction and the evolved interaction, respectively, we can extract a unique set of central and tensor correlation functions. After multiplying the mapping equation \eqref{eq:srgcorr_definition} with $\bra{r(LS)JT}$ and $\bra{r(L'S)JT}$, respectively, and using Eqs. \eqref{eq:corr_tensor_states2} and \eqref{eq:ucom_central_states}, we obtain a system of two equations
\eqmulti{ \label{eq:srgcorr_coupledpw}
  &\begin{pmatrix}
  \phi^{(0)}_{L}(r) \\
  \phi^{(0)}_{L'}(r)
  \end{pmatrix}
  =
  \frac{\Rm(r)}{r} \sqrt{\DRm(r)} \times \\
  &\qquad\times
  \begin{pmatrix}
  \cos\theta_J(r) && \sin\theta_J(r) \vphantom{\phi^{(\alpha)}_{L}} \\
  -\sin\theta_J(r) && \cos\theta_J(r) \vphantom{\phi^{(\alpha)}_{L'}}
  \end{pmatrix}
  \begin{pmatrix}
  \phi^{(\alpha)}_{L}(\Rm(r)) \\
  \phi^{(\alpha)}_{L'}(\Rm(r))
  \end{pmatrix} \;,
}
from which the correlation functions  $\Rm(r)$ and $\vartheta(r)$ can be determined.

Since the central correlation function acts on both orbital components in the same way and since the transformation matrix in \eqref{eq:srgcorr_coupledpw} is unitary, we can determine $\Rm(r)$ without knowing $\vartheta(r)$. By considering the sum of the squares of the two orbital components we obtain from \eqref{eq:srgcorr_coupledpw} the identity
\eqmulti{
  &[\phi^{(0)}_{L}(r)]^2 + [\phi^{(0)}_{L'}(r)]^2 
  = \frac{[\Rm(r)]^2}{r^2} \DRm(r)\; \times \\
  &\qquad\times \big( [\phi^{(\alpha)}_{L}(\Rm(r))]^2 + [\phi^{(\alpha)}_{L'}(\Rm(r))]^2 \big) \;.
}
which corresponds to \eqref{eq:srgcorr_noncoupledpw_pre} for the non-coupled case. The correlation function $\Rm(r)$ can then be determined iteratively from the integral equation
\eq{
  [\Rm(r)]^3 
  = 3 \int_0^r d\xi\; \xi^2 \frac{[\phi^{(0)}_{L}(\xi)]^2 + [\phi^{(0)}_{L'}(\xi)]^2} 
    {[\phi^{(\alpha)}_{L}(\Rm(\xi))]^2 + [\phi^{(\alpha)}_{L'}(\Rm(\xi))]^2} \;.
}
Once $\Rm(r)$ is known, the system \eqref{eq:srgcorr_coupledpw} reduces to a set of two nonlinear equations for $\theta_J(r) = 3 \sqrt{J(J+1)}\; \vartheta(r)$, which can be solved numerically for each $r$.  Eventually, we obtain discretized correlation functions $\Rm(r)$ and $\vartheta(r)$ also for the coupled partial waves.

\subsection{SRG-Generated Correlation Functions for the AV18}

We use this mapping scheme to determine a set of correlation functions for the AV18 potential. In line with the previous applications of the UCOM approach we allow for different correlation functions in the different spin-isospin-channels. An explicit angular-momentum dependence of the correlation functions is not included. Therefore, the lowest partial wave for each spin-isopin-channel is used to fix the correlation functions.

\begin{figure}
\includegraphics[width=0.8\columnwidth]{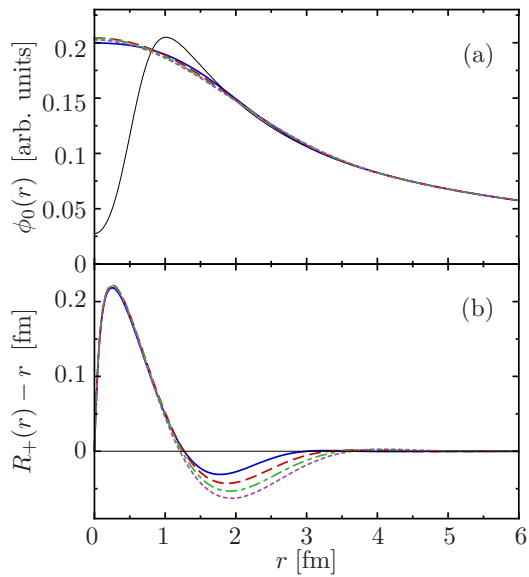}
\caption{(color online) Radial wavefunction $\phi_0(r)$ and resulting correlation function $\Rp(r)-r$ for the $^1S_0$ partial wave for different flow parameters: $\bar{\alpha}=0.02\,\fm^4$ (\linethinsolid[FGBlue]), $0.04\,\fm^4$ (\linethindashed[FGRed]), $0.06\,\fm^4$ (\linethindashdot[FGGreen]), $0.08\,\fm^4$ (\linethindotted[FGViolet]). The thin solid curve in panel (a) shows the corresponding wavefunction for the initial potential used for the mapping.}
\label{fig:srgcorr_corr_ST01}
\end{figure}

As an example for a non-coupled channel, we discuss the $^1S_0$ partial wave. In this partial wave the potential does not support a bound state, so all discrete eigenstates correspond to continuum states with a discretization resulting from the boundary conditions employed for the numerical solution. The correlation function $\Rm(r)$ obtained from \eqref{eq:srgcorr_noncoupledpw} is inverted numerically in order to provide the correlation function $\Rp(r)$, which is used in all subsequent calculations. Figure \ref{fig:srgcorr_corr_ST01} depicts the radial wavefunctions of the lowest eigenstate for different values of the flow parameter $\bar{\alpha}$ as well as the correlation functions $\Rp(r)$ resulting from the mapping. 

The shape of the correlation functions $\Rp(r)$ is characteristic and can be understood intuitively in terms of a coordinate transformation as mentioned in Sec. \ref{sec:ucom_states}. At short distances the quantity $\Rp(r)-r$, which can be viewed as a radial shift distance, is positive. Thus, keeping the transformation \eqref{eq:ucom_central_states} in mind, probability amplitude is shifted from small towards larger relative distances. At some distance, $\Rp(r)-r$ changes sign and becomes negative, corresponding to a shift towards smaller $r$. The change of sign appears right within the most attractive region of the potential, i.e. the probability amplitude is concentrated there. It is worthwhile noting, that all correlation functions automatically have finite range, which warrants that initial and transformed potential are phase-shift equivalent.

For the different SRG parameters $\bar{\alpha}$ used in Fig. \ref{fig:srgcorr_corr_ST01} the short-range part of the correlation function $\Rp(r)$ does not change. These short-range and high-momentum correlations are removed in the initial stages of the SRG evolution and are unaffected by the further evolution (also see Fig. \ref{fig:srg_illu}). 
Only the long-range part of the correlation functions depends on the flow parameter---with increasing $\bar{\alpha}$ correlations of longer and longer range are removed through the SRG transformation leading to correlation functions of increasing range. Thus, the SRG parameter $\bar{\alpha}$ and the range of the UCOM correlation functions are directly connected. 

\begin{figure}
\includegraphics[width=0.8\columnwidth]{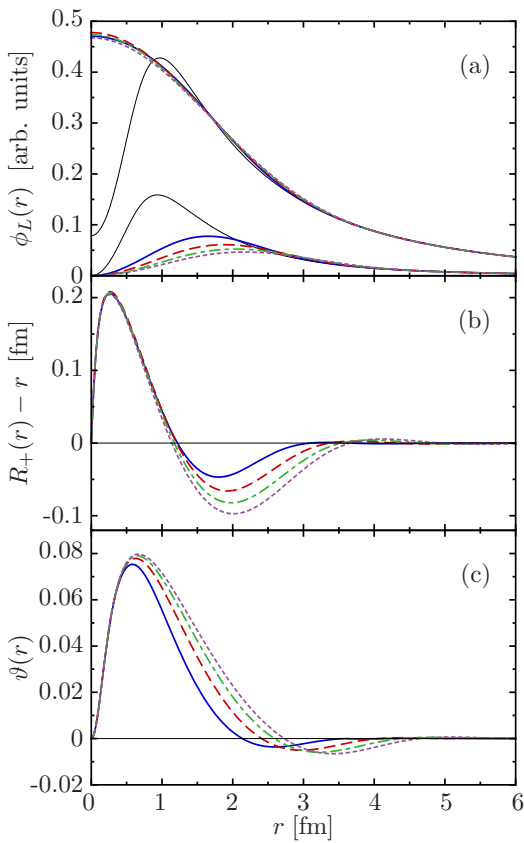}
\caption{(color online) Radial wavefunctions of the $S$ and $D$-wave component, $\phi_0(r)$ and $\phi_2(r)$ and resulting central correlation function $\Rp(r)-r$ as well as the tensor correlation function $\vartheta(r)$ for the $^3S_1-{}^3D_1$ partial waves for different flow parameters: $\bar{\alpha}=0.02\,\fm^4$ (\linethinsolid[FGBlue]), $0.04\,\fm^4$ (\linethindashed[FGRed]), $0.06\,\fm^4$ (\linethindashdot[FGGreen]), $0.08\,\fm^4$ (\linethindotted[FGViolet]). The thin solid curve in panel (a) shows the corresponding wavefunctions for the initial potential used for the mapping.}
\label{fig:srgcorr_corr_ST10}
\end{figure}

As an example for a coupled channel, we consider the $^3S_1-{}^3D_1$ partial waves and the two-body ground state of the deuteron. From the S and D-wave component of the wave function the central and tensor correlation functions are extracted by solving Eq. \eqref{eq:srgcorr_coupledpw}. The input wavefunctions and the resulting correlation functions $\Rp(r)$ and $\vartheta(r)$ are depicted in Fig. \ref{fig:srgcorr_corr_ST10}. The central correlation functions show the same structure as in the $^1S_0$ channel with a short-range component independent of the flow parameter. The tensor correlation functions $\vartheta(r)$ also exhibit positive and negative contributions, with a dominant positive section at short ranges. The dependence on $\bar{\alpha}$ is stronger than for the central correlation function. With increasing $\bar{\alpha}$ the decreasing slope of $\vartheta(r)$ as a whole is shifted towards larger $r$ which also effects the behavior around $r\approx1\,\fm$. The tensor correlations do not show the clear separation of short- and long-range effects that was observed for the central correlations.    

\begin{figure}
\includegraphics[width=0.8\columnwidth]{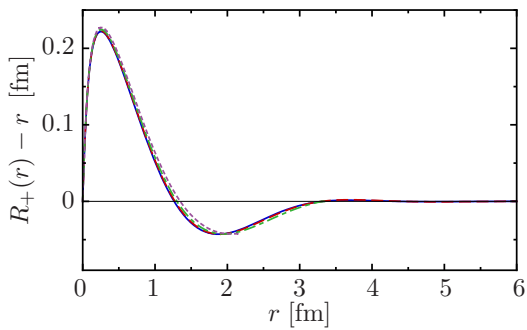}
\caption{(color online) Central correlation functions $\Rp(r)-r$ derived from the ground state (\linethinsolid[FGBlue]), the second (\linethindashed[FGRed]), the fourth (\linethindashdot[FGGreen]), and the sixth excited states  (\linethindotted[FGViolet]) in the $^1S_0$ partial wave for the flow parameter $\bar{\alpha}=0.04\,\fm^4$. The correlation functions are plotted only up to the position of the first zero of the radial wavefunctions.}
\label{fig:srgcorr_excited_ST01}
\end{figure}

So far, only the two-body ground state for the given partial wave has been used to extract the correlation functions. In principle, any other state of the two-body spectrum can be used as well. It is therefore important to check the sensitivity of the resulting correlation functions on the choice of the eigenstate. In Fig. \ref{fig:srgcorr_excited_ST01} we report the correlation functions $\Rp(r)$ extracted from four different $^1S_0$ eigenstates spanning a range of two-body energies from $0$ to $20\,$MeV. The correlation functions are surprisingly stable in this energy range, showing only a slight tendency towards longer-ranged correlators for larger excitation energies. The same holds true for the other partial waves and the tensor correlation functions. This is another indication that the generator of the UCOM method encapsulates the relevant physics of short-range correlations in a simple explicit operator transformation. In the following we will always use the lowest state of the two-body spectrum to fix the UCOM correlation functions.

This construction can be repeated for each partial wave leading to a different set of correlation functions for each combination of angular momenta, spin, and isospin. In the standard UCOM framework we restrict ourselves to a set of correlation functions depending on spin and isospin alone, i.e. there are four different central correlation functions $\Rp(r)$ for the different combinations of $S=0,1$ and $T=0,1$ and two different tensor correlation functions $\vartheta(r)$ for $S=1$ and $T=0,1$. They are determined from a mapping in the lowest partial wave for each $S$ and $T$. As a result of this restriction the UCOM transformation is not specifically optimized for the higher partial waves. However, since the centrifugal barrier suppresses the short-range part of the relative wavefunctions in higher partial waves, the impact of short-range correlations is reduced in any case. Eventually, any residual correlations not covered explicitly by the UCOM transformation have to be described by the many-body method, which uses the UCOM interactions as input.

\subsection{Comparison with Variationally-Optimized Correlators}

The set of SRG-generated UCOM correlation functions can be compared to the correlation functions used in previous UCOM calculations. Those were extracted within a variational scheme using simple parametrizations of the correlation functions $\Rp(r)$ and $\vartheta(r)$, whose parameters were determined from a minimization of a single momentum space matrix element of the correlated interaction---the diagonal $q=q'=0$ matrix element. In this approach the tensor correlation functions are subject to a constraint on the range defined via the volume integral 
\eq{
  I_{\vartheta} 
  = \int dr\; r^2 \vartheta(r)
}
in order to isolate the short-range component of the two-body correlations. A detailed discussion of this scheme including the parametrizations and optimal parameters can be found in Ref. \cite{RoHe05}. There is a correspondence between the range parameter $I_{\vartheta}$ and the flow parameter $\alpha$. Both define the separation scale between short-range and long-range correlations, where the short range component is eliminated by the UCOM or SRG-transformation. For the variationally optimized correlators this separation scale is mainly determined by the range of the tensor correlators, which is controlled directly via the integral constraint $I_{\vartheta}$.

\begin{figure}
\includegraphics[width=1\columnwidth]{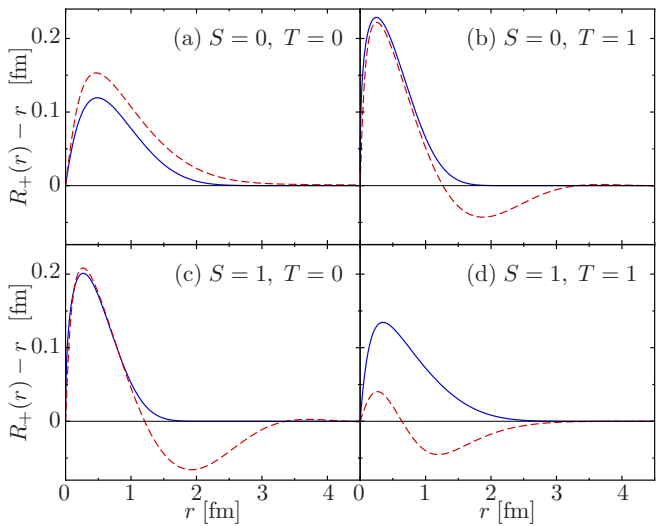}
\caption{(color online) Central correlation functions $\Rp(r)-r$ for different spin and isospin channels obtained from an energy minimization (\linethinsolid[FGBlue]) and from the SRG-mapping (\linethindashed[FGRed]). }
\label{fig:srgcorr_comparison_Rp}
\end{figure}

\begin{figure}
\includegraphics[width=1\columnwidth]{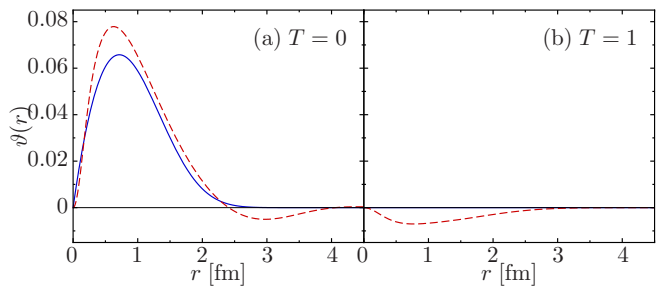}
\caption{(color online) Tensor correlation functions $\vartheta(r)$ obtained from an energy minimization (\linethinsolid[FGBlue]) and from the SRG-mapping (\linethindashed[FGRed]). }
\label{fig:srgcorr_cpmparison_Th}
\end{figure}

\begin{figure*}
\includegraphics[width=0.5\textwidth]{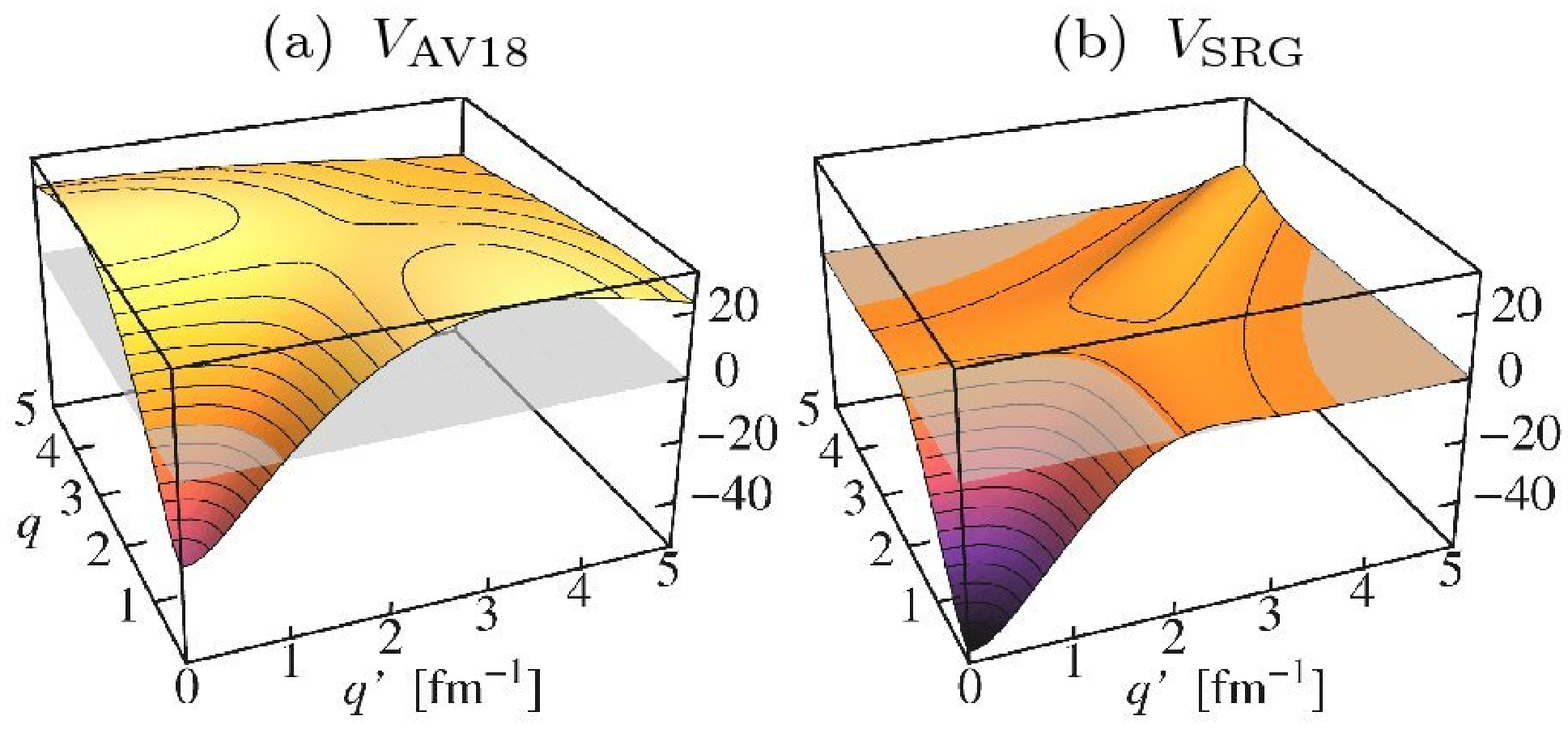}%
\includegraphics[width=0.5\textwidth]{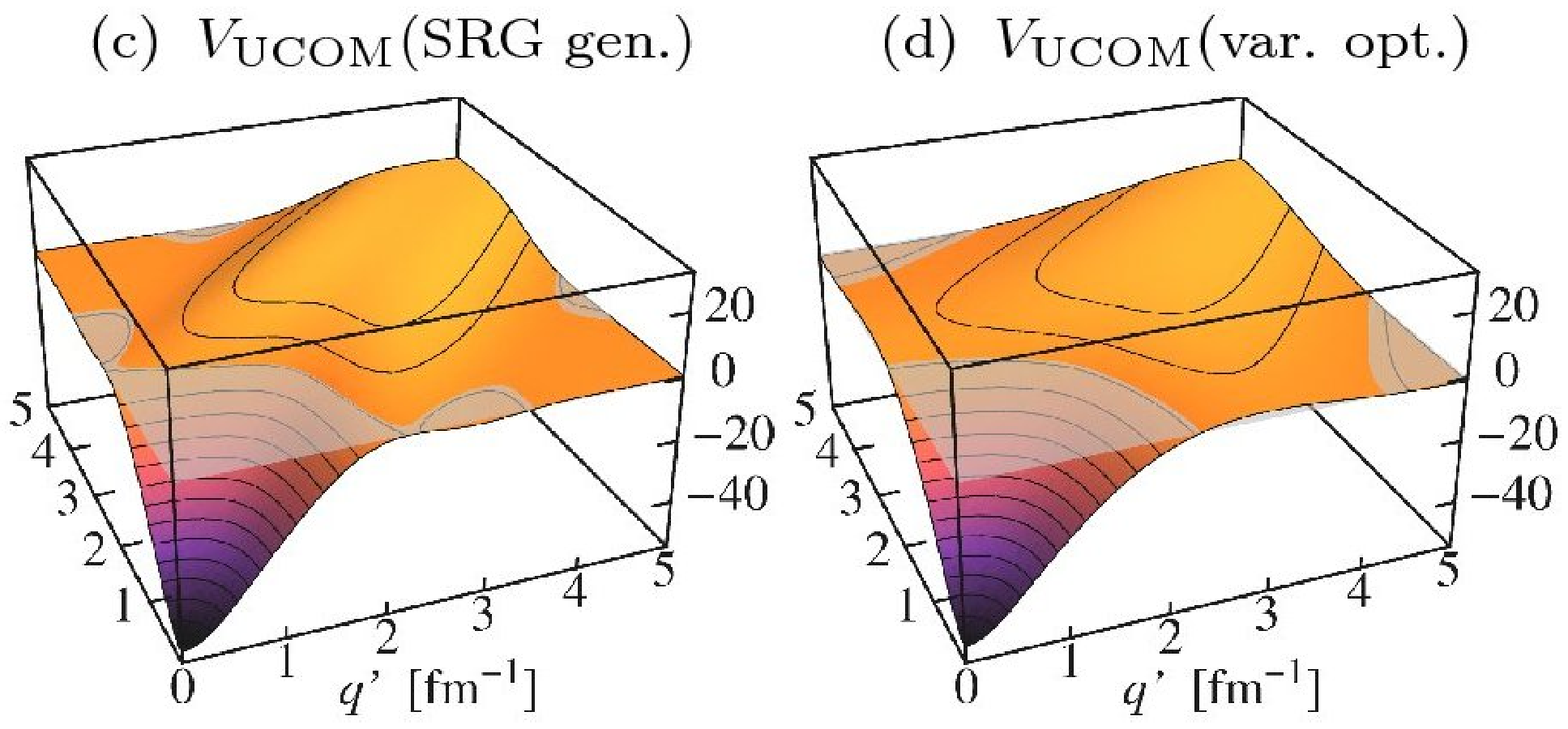}
\caption{(color online) Momentum-space matrix elements $\matrixe{q(LS)JT}{\circ}{q'(L'S)JT}$ (in units of $\MeV\fm^3$) for the $^1S_0$ partial wave obtained from (a) the initial AV18 potential, (b) the SRG-evolved interaction ($\bar{\alpha}=0.03\,\fm^4$), (c) the UCOM-transformed interaction using the SRG-generated correlators ($\bar{\alpha}=0.04\,\fm^4$), and (d) the UCOM-transformed interaction using the variationally optimized correlators.}
\label{fig:srgcorr_meq}
\end{figure*}
\begin{figure*}
\includegraphics[width=0.5\textwidth]{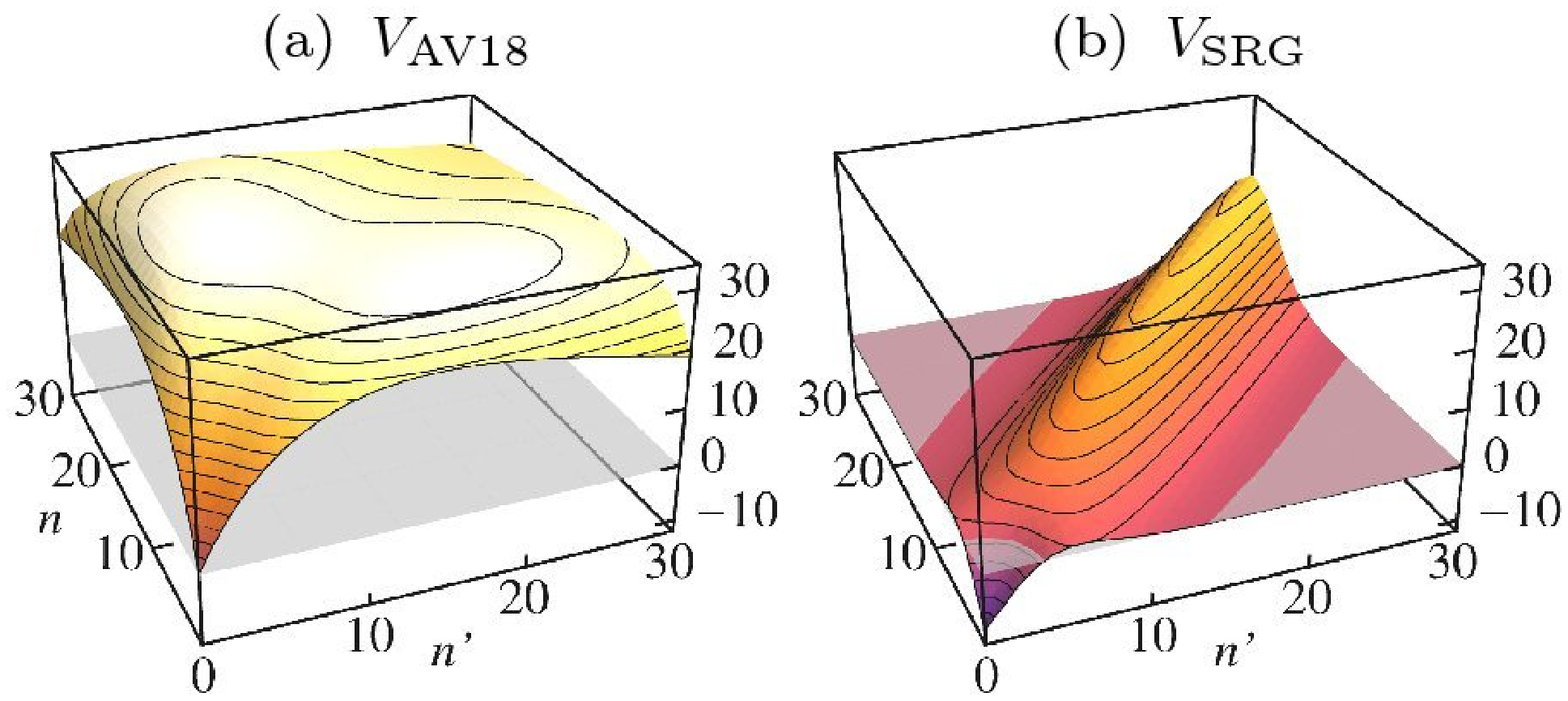}%
\includegraphics[width=0.5\textwidth]{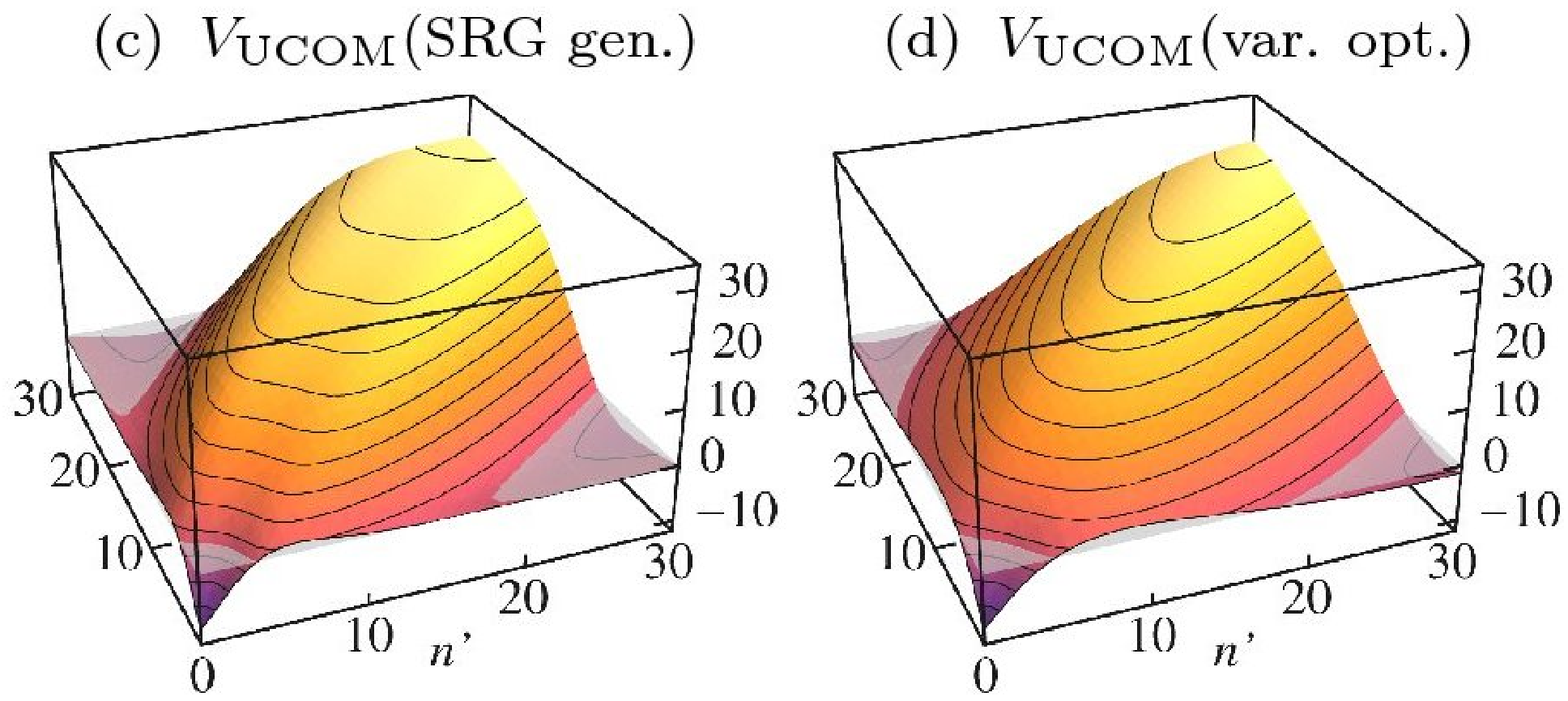}
\caption{(color online) Relative harmonic-oscillator matrix elements $\matrixe{n(LS)JT}{\circ}{n'(L'S)JT}$ (in units of $\MeV$) for an oscillator frequency $\hbar\Omega=20\,\MeV$ in the $^1S_0$ partial wave obtained with the same interactions as in Fig. \ref{fig:srgcorr_meq}.}
\label{fig:srgcorr_meho}
\end{figure*}

In Figs. \ref{fig:srgcorr_comparison_Rp} and \ref{fig:srgcorr_cpmparison_Th} the central and tensor correlation functions, respectively, resulting from energy minimization and the SRG-mapping are compared. The range constraint $I_{\vartheta,\text{even}}=0.09\,\fm^3$ and the flow parameter $\bar{\alpha}=0.04\,\fm^4$, respectively, are chosen such that the binding energy of \elem{He}{4} resulting from a converged no-core shell model calculation is in agreement with experiment (cf. Sec. \ref{sec:ncsm}). In the dominant even channels the short-range behavior of the correlation functions $\Rp(r)$ agrees very well, as depicted in Fig. \ref{fig:srgcorr_comparison_Rp}. The long-range behavior of the SRG-generated correlators is dominated by the negative section in $\Rp(r)-r$, which was not considered in the parametrizations used for the previous determination. For the triplet-even tensor correlator, the shape of the correlation functions is slightly different, but the gross behavior agrees, as seen from Fig. \ref{fig:srgcorr_cpmparison_Th}. The deviations in the singlet-odd channel result from an additional range-constraint imposed for the variational construction of $\Rp(r)$, since the interaction is purely repulsive in this channel. In the triplet-odd channel the tensor correlation function was switched-off for the variational determinantion, i.e. $I_{\vartheta,\text{odd}}=0\,\fm^3$, which also induces a different central correlation function.  

Aside from these small quantitative differences, the SRG-mapping is conceptionally superior to the variational construction. There is a single well-defined parameter, the flow parameter $\bar{\alpha}$, that unambiguously spans a family of correlation functions. For the variational optimization, one always has the freedom to choose different parametrizations and different ways to constrain the correlator ranges, which complicates a consistent and unambiguous treatment.

\subsection{Comparison of Matrix Elements}
\label{sec:srgcorr_me}

It is important to realize that the UCOM-transformation using SRG-generated correlators is not equivalent to a direct SRG-evolution. For the construction of the UCOM correlators only a single eigenstate from the low-energy part of the two-body spectrum of the SRG-evolved interaction is used. This contains the essential information on the decoupling of  low-momentum from high-momentum modes. However, this does not guarantee a decoupling among high-momentum modes.  

This difference is illustrated in Fig. \ref{fig:srgcorr_meq} using momentum-space matrix elements $\matrixe{q(LS)JT}{\circ}{q'(L'S)JT}$ for the $^1S_0$ partial wave. Here and in the following comparisons of SRG-evolved and UCOM-transformed interactions we fix the parameters such that the \elem{He}{4} binding energy obtained in the no-core shell model for each of the transformed interactions is in agreement with experiment. For the SRG-evolved interaction this leads to $\bar{\alpha}=0.03\,\fm^4$, for UCOM with SRG-generated correlators we obtain $\bar{\alpha}=0.04\,\fm^4$, and for UCOM with variationally optimized correlators we use $I_{\vartheta,\text{even}}=0.09\,\fm^3$. The momentum-space matrix elements of the UCOM-transformed interaction are computed using the analytic form discussed in \cite{RoHe05}.

In comparison to the initial AV18 interaction, all the unitarily transformed interactions exhibit a strong reduction of the off-diagonal matrix elements and an enhancement of the low-momentum sector. In the high-momentum regime, the SRG-evolved interaction by construction shows a narrow band-diagonal structure, i.e. the decoupling is effective at all momenta. The UCOM-transformed interactions for both, the SRG-generated correlators and the ones determined variationally, exhibit larger off-diagonal contributions connecting different high-momentum states.  The SRG-generated correlators lead to a broad band of non-vanishing matrix elements along the diagonal at high momenta---much broader than for the SRG-evolved interaction.     

The difference is even more pronounced when going from momentum-space to the harmonic oscillator basis that will be used for the following many-body calculations. In Fig. \ref{fig:srgcorr_meho} we present the relative harmonic-oscillator matrix elements $\matrixe{n(LS)JT}{\circ}{n'(L'S)JT}$ for the same $^1S_0$ partial wave and the same interactions. The UCOM matrix elements are computed directly in the harmonic oscillator basis following Ref. \cite{RoHe05}, whereas the SRG matrix elements result from an evolution in momentum space and a subsequent transformation into the oscillator basis. Although the SRG-evolution is tailored for a pre-diagonalization in momentum space, the harmonic oscillator matrix elements also show a narrow band-diagonal structure for all radial quantum numbers $n$. The UCOM-transformed interactions have more off-diagonal contributions at large $n$ and thus a stronger coupling between high-lying states. However, the behavior at small $n$ and the decoupling of states with small $n$ from those with large $n$ is very similar.

The matrix elements show that UCOM is as efficient as SRG in decoupling low- and high-lying states, but has a different structure in the high-$q$ or large-$n$ regime. The former is most relevant for the convergence properties of the interaction, the latter influences the behavior when going to heavier systems.

\section{Few- \& Many-Body Calculations}
\label{sec:ncsmhf}

As a first application and test of the SRG-generated UCOM correlators we discuss No-Core Shell Model (NCSM) calculations for the ground states of \elem{H}{3} and \elem{He}{4} and Hartree-Fock calculations for heavier closed-shell nuclei. These calculations shed light on the similarities of and differences between SRG and UCOM-transformed interactions relevant for nuclear structure.

\subsection{No-Core Shell Model for \elem{H}{3} and \elem{He}{4}}
\label{sec:ncsm}

\begin{figure*}
\includegraphics[width=1\textwidth]{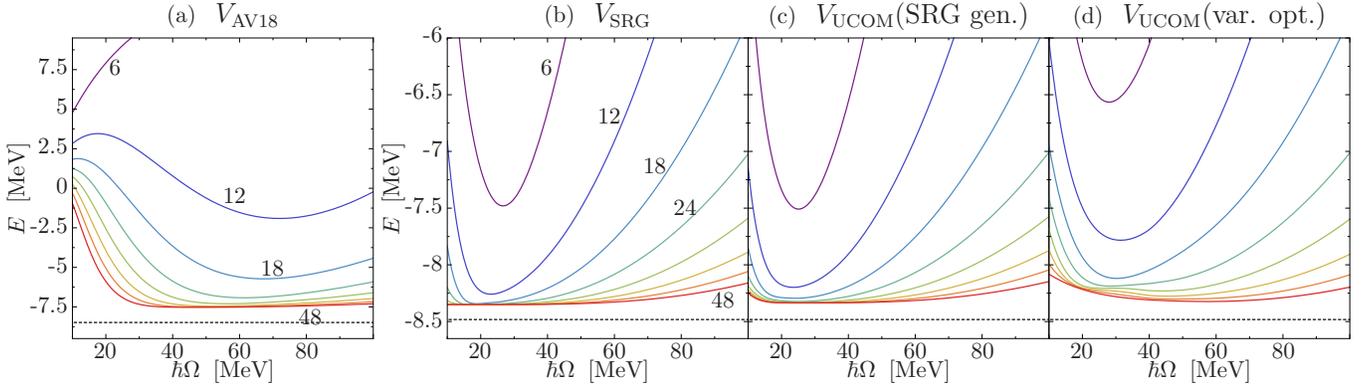}
\caption{(color online) Convergence behavior of the ground-state energy of \elem{H}{3} obtained in no-core shell model calculations as function of $\hbar\Omega$ for different potentials. The various curves correspond to different $N_{\max}\hbar\Omega$ model spaces in the range $N_{\max}=6$, $12$, $18$,..., $48$ as indicated by the labels. The different potentials are: (a) untransformed AV18 potential (note the different energy scale). (b) SRG-evolved AV18 potential for $\bar{\alpha}=0.03\,\fm^4$. (c) UCOM-transformed AV18 potential using the SRG-generated correlators for $\bar{\alpha}=0.04\,\fm^4$. (d) UCOM-transformed AV18 potential using the standard correlators constructed by energy minimization. The dashed horizontal line indicates the experimental ground-state energy.}
\label{fig:ncsm_H3}
\end{figure*}
\begin{figure*}
\includegraphics[width=1\textwidth]{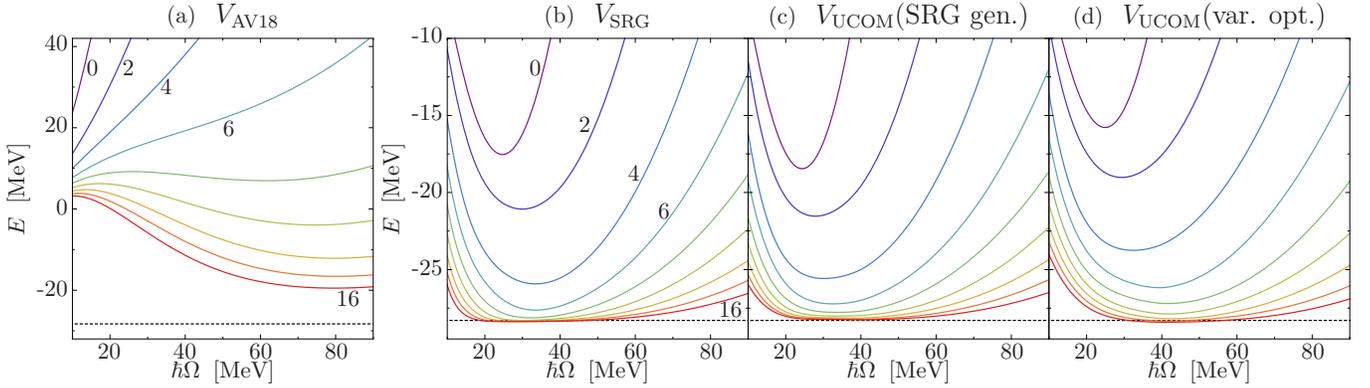}
\caption{(color online) Convergence behavior of the ground-state energy of \elem{He}{4} obtained in no-core shell model calculations as function of $\hbar\Omega$ for different potentials. The various curves correspond to different $N_{\max}\hbar\Omega$ model spaces in the range $N_{\max}=0$, $2$, $4$,..., $16$ as indicated by the labels. The different potentials are as described in Fig. \ref{fig:ncsm_H3}.}
\label{fig:ncsm_He4}
\end{figure*}

For light nuclei the NCSM provides detailed insight into the convergence behavior of the different interactions and thus allows to disentangle the effects of short- and long-range correlations. For a given $N_{\max}\hbar\Omega$ model space of the NCSM we diagonalize the translationally invariant many-body Hamiltonian consisting of intrinsic kinetic energy and two-body SRG- or UCOM-transformed interaction directly. No additional Lee-Suzuki transformation is employed. All calculations were performed with the Jacobi-coordinate NCSM code of P. Navr\'atil \cite{NaKa00}.

We will use only the two-body terms of the transformed interactions in these calculations and discard three-body and higher-order contributions that are inevitably generated by the unitary transformation. We are thus treating the transformed two-body terms as a new realistic interaction, which is phase-shift equivalent to the initial AV18 potential. The energy eigenvalues obtained with these two-body interactions in a many-body system are different from the eigenvalues of the initial potential---only if all many-body terms of the transformed Hamiltonian were included, unitarity would guarantee invariance of the eigenvalues. As pointed out in Refs. \cite{RoHe05,RoNa07,Roth08} we can use this fact to choose a unitary transformation that requires minimal many-body forces for the description of the ground-state energy of a selected nucleus. For the following discussion we fix the parameter controlling the SRG or UCOM-transformation such that the converged NCSM ground-state energy of \elem{He}{4} is in agreement with experiment, i.e. we minimize the contribution of three- and four-body interactions to the \elem{He}{4} ground-state energy. As mentioned earlier, this condition is fulfilled for the SRG-evolved potential with $\bar{\alpha}=0.03\,\fm^4$ \cite{BoFu07b}, the UCOM-transformed potential using the SRG-generated correlation functions with $\bar{\alpha}=0.04\,\fm^4$, and the UCOM-transformed potential using the variationally optimized correlation functions with $I_{\vartheta,\text{even}}=0.09\,\fm^3$. 

In Figs. \ref{fig:ncsm_H3} and \ref{fig:ncsm_He4} we present the ground-state energies of \elem{H}{3} and \elem{He}{4}, respectively, as function of the oscillator frequency $\hbar\Omega$ for different model space sizes $N_{\max}$. First of all, the systematics of the $N_{\max}$-dependence reveals the huge difference in the convergence behavior of the initial AV18 interaction and the different transformed interactions (note the different energy scales). The transformed interactions lead to a self-bound ground state within a $0\hbar\Omega$ model space already. For \elem{He}{4} the  $0\hbar\Omega$  space consists of a single Slater determinant which cannot describe any correlations. Thus the large change of the $0\hbar\Omega$ energy proves that the unitary transformations have eliminated the components of the AV18 interaction that induce short-range correlations in the many-body states. These are the components that generate large matrix elements far-off the diagonal which couple low-lying and high-lying basis states. The residual correlations resulting from near-diagonal matrix elements can be described in model spaces of moderate size. Hence the transformed interactions show a rapid convergence---for \elem{He}{4} at $N_{\max}\gtrsim 10$---where the initial AV18 is still far from the converged result. 

The comparison of the results obtained with the three transformed interactions reveals a few subtle but important differences. The convergence properties of the SRG-evolved interaction and the UCOM-transformed interaction for the SRG-generated correlators, shown in Figs. \ref{fig:ncsm_H3}(b) and (c) and in Figs. \ref{fig:ncsm_He4}(b) and (c), is very similar. Both exhibit a very regular convergence pattern. As function of the oscillator frequency there is a single minimum, which flattens rapidly with increasing $N_{\max}$ leading to a converged ground-state energy which is constant over an extended range of frequencies $\hbar\Omega$. These similarities indicate that the obvious differences of the matrix elements in the high-$q$ or large-$n$ sector, as discussed in Sec. \ref{sec:srgcorr_me}, are irrelevant for the convergence in light nuclei.

The two UCOM-transformed interactions, using the SRG-generated and the variationally optimized correlation functions, respectively, show slightly different convergence patterns as seen in  Figs. \ref{fig:ncsm_H3}(c) and (d) and in Figs. \ref{fig:ncsm_He4}(c) and (d). Overall, the SRG-generated correlators lead to lower energies in small model spaces and to a faster and more regular convergence. For \elem{H}{3} in particular, the variationally optimized correlators develop a double-minimum structure for model spaces around $N_{\max}=24$, which disturbs the smooth convergence. Eventually, at large $N_{\max}$ the minimum shifts to large values of $\hbar\Omega$. The appearance of the second minimum indicates that the correlation functions do not describe certain features of the interparticle correlation properly. The NCSM corrects for these deficiencies as soon as the model space is sufficiently large to resolve the relevant length scales. If the defects are well localized with respect to the interparticle distance, then huge model spaces are required to resolve them. The fact that the SRG-generated correlators work much better, can be traced back to the negative sections in the correlation functions $\Rp(r)-r$, which pull in probability amplitude from larger interparticle distances into the attractive region of the interaction. This localized modification of the two-body density can be described in the NCSM only with large model spaces, leading to a change in the convergence pattern.

\subsection{Hartree-Fock for Heavier Nuclei}

In order to highlight the differences among the transformed interactions emerging in heavier nuclei, we present simple Hartree-Fock (HF) calculations for selected nuclei with closed $j$-shells throughout the nuclear chart. We use the HF implementation discussed in detail in Ref. \cite{RoPa06} based on the intrinsic Hamiltonian $\HO_{\text{int}}= \TO-\TO_{\cm} + \VO$ including all charge dependent and electromagnetic terms of the SRG- and UCOM-transformed AV18 potential. The single-particle states are expanded in the harmonic oscillator basis with an oscillator parameter selected via a minimization of the HF energy. All calculations were performed with a basis including 13 major oscillator shells which warrants convergence of the HF ground-state energies. 

Of course, the HF many-body state, being a Slater determinant, cannot describe any correlations by itself. Thus compared to the NCSM calculation in the previous section, HF can only provide results at the level of a $0\hbar\Omega$ space. The energy gain observed in the NCSM by increasing the size of the model space resulting from residual correlations cannot be obtained in HF. To recover the effect of these residual correlations, extensions beyond HF, e.g. in the framework of many-body perturbation theory, have to be considered. Nonetheless, the HF solution provides valueable information on the systematics of ground-state energies and rms-radii. For the following conclusions we solely rely on the fact that the HF energies provide a variational upper bound for the exact ground-state energies and that residual correlations, in the case of the UCOM interactions, change the binding energies per particle by an almost constant amount \cite{RoPa06}. 

\begin{figure}
\includegraphics[width=1\columnwidth]{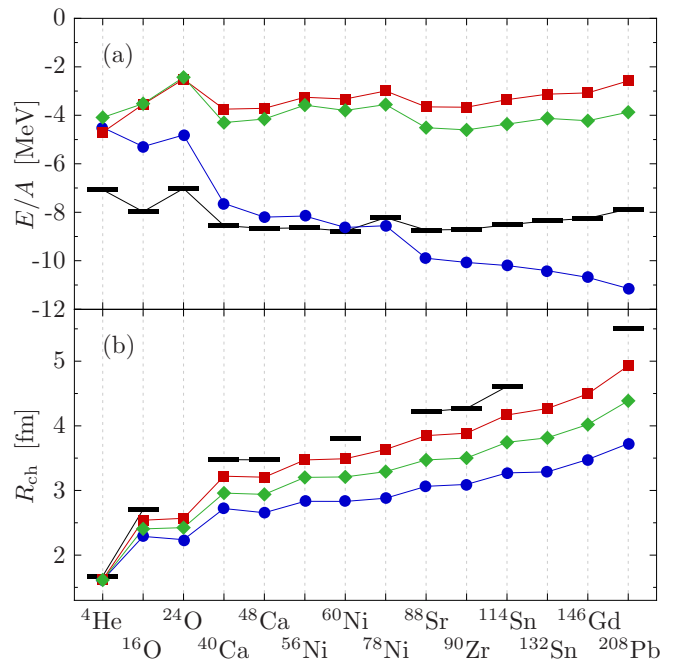}
\caption{(color online) Hartree-Fock ground-state energies per particle and charge radii obtained with the SRG-evolved AV18 potential (\symbolcircle[FGBlue]), the UCOM-transformed AV18 using SRG-generated correlators (\symbolbox[FGRed]), and the UCOM-transformed interaction using the variationally optimized correlators (\symboldiamond[FGGreen]). The parameters of all transformed interactions are the same as in the NCSM calculations. Experimental data is represented by black bars \cite{AuWa95,VrJa87}.}
\label{fig:hf}
\end{figure}

In Fig. \ref{fig:hf} we summarize the HF results for ground-state energies and charge rms-radii for a range of nuclei from \elem{He}{4} to \elem{Pb}{208}. We use the same transformed interactions as for the NCSM calculations: the SRG-evolved interaction ($\bar{\alpha}=0.03\,\fm^4$), the UCOM-transformed interaction using SRG-generated correlators  ($\bar{\alpha}=0.04\,\fm^4$), and the UCOM-tranformed interaction using variationally optimized correlators ($I_{\vartheta,\text{even}}=0.09\,\fm^3$), all derived from the AV18 potential.

A systematic difference is observed between the SRG-evolved and the UCOM-transformed interactions. In the case of the UCOM-transformed interactions, the energies per nucleon are almost constant as function of mass number $A$. They resemble the systematics of the experimental binding energies up to a constant energy shift. The inclusion of the effect of residual correlations on the energy, e.g. by means of many-body perturbation theory, will shift the HF energies right into the region of the experimental data as was demonstrated in Ref. \cite{RoPa06} for the UCOM interaction with variationally optimized correlators. In contrast, the SRG-evolved interaction shows a systematic linear increase of the binding energy per nucleon with $A$ \cite{BoFu07b}, leading to a dramatic overbinding for heavier nuclei already on the HF level. The inclusion of correlations beyond HF will lead to even more drastic deviations. 

The charge radii also reflect this difference. The SRG-evolved interaction predicts very small charge radii for heavy nuclei, for \elem{Pb}{208} its about $1.8\,\fm$ smaller than the experimental value. The two UCOM-transformed interactions generate radii which are also too small, but much closer to the experimental values. It is interesting to note that the difference in the radii predicted with the UCOM interactions is rather large, those obtained with the SRG-generated correlators are significantly closer to experiment.   

One can interpret the different systematics in terms of the impact of three-body interactions which have been omitted here. For the SRG-transformed interaction there is a clear need for a strongly repulsive three-body interaction. Given the huge effect on binding energies and charge radii the structure of the states will be changed completely by the additional three-body force. For the UCOM-transformed interactions a supplementary three-body force will have a much smaller effect. One might expect a small correction to the systematics of the charge radii due to three-body forces, for the energies the systematics is already reproduced by the two-body force.

\section{Conclusions}

The Unitary Correlation Operator Method and the Similarity Renormalization Group are two methods to tackle short-range correlations in the nuclear many-body problem by means of unitary transformations. Though both methods start from a different conceptual background---coordinate-space picture of short-range correlations and pre-diagonalization via a flow evolution, respectively---both lead to a decoupling of low-momentum and high-momentum modes. Moreover, there are certain formal connections, e.g. regarding the initial structure of the generators, and we have shown how to use the SRG-scheme to construct correlation functions for the UCOM transformation by means of a mapping of two-body eigenstates. These SRG-generated UCOM correlation functions provide an alternative to the previous correlation functions obtained by a variational procedure and will be explored further. 

The resulting phase-shift equivalent effective interactions show similarities but also differences. Because of the decoupling of low-$q$ or small-$n$ states from high-$q$ or large-$n$ states, both the SRG-evolved and the UCOM-transformed interactions lead to a rapid convergence of NCSM calculations for light nuclei. However, the behavior of matrix elements in the high-$q$ or large-$n$ sector is quite different. The SRG-evolution causes a pre-diagonalization at all momentum scales, i.e. it also leads to a decoupling among the high-$q$ or large-$n$ states. The UCOM-transformed interaction generates a stronger coupling among high-lying states, i.e. the prediagonalization in the high-$q$ or large-$n$ regime is not as perfect. This difference, together with the independence of the UCOM transformation on angular momentum, seems crucial when going to heavier systems. Simple HF calculations with a pure two-body UCOM-transformed and SRG-evolved interactions reveal a different systematic behavior of the binding energies as function of mass number. The SRG-interactions lead to a systematic overbinding for heavier nuclei already at the HF level, whereas the UCOM-transformed interaction result in a constant energy per particle. From this observation one might conclude that three-body interactions have to have a large effect in the SRG-scheme, whereas their impact in the UCOM-framework is much smaller.

\section*{Acknowledgments}

This work is supported by the Deutsche Forschungsgemeinschaft through contract SFB 634 and through the GSI F\&E program. 


\end{document}